\documentclass{aa}
\usepackage[varg]{txfonts}
\usepackage{ifthen}
\usepackage{url}
\usepackage{amsmath}
\usepackage{bm}
\usepackage{lscape}
\usepackage{rotating}
\usepackage{graphicx}
\usepackage{supertabular}
\usepackage{pdfpages}
\usepackage{footnote}
\usepackage{amssymb}
\usepackage{multirow}
\usepackage{xcolor}
\usepackage{subfigure}
\usepackage{hyperref} 

\definecolor{myurlcolor}{HTML}{0000FF}

\hypersetup{
  colorlinks   = true,    
  urlcolor     = myurlcolor,   
  linkcolor    = myurlcolor,   
  citecolor    = myurlcolor      
}

\begin{document}

\title{Revisiting the proper motions of M31 and M33 using massive supergiant stars with Gaia DR3}

\author{Hao Wu \inst{1, 2}, Yang Huang \inst{3,4,5}, Huawei Zhang \inst{1,2,5}, Qikang Feng \inst{1, 2}}

\institute{Department of Astronomy, School of Physics, Peking University, Beijing 100871, China; \\ \email{zhanghw@pku.edu.cn} \and Kavli Institute for Astronomy and Astrophysics, Peking University, Beijing 100871, China \and School of Astronomy and Space Science, University of Chinese Academy of Science, Beijing 100049, China; \\ \email{huangyang@ucas.ac.cn} \and National Astronomical Observatories, Chinese Academy of Sciences, Beijing 100101, China \and Corresponding authors}

\titlerunning{The proper motion of M31 and M33}
\authorrunning{Wu et al.}

\abstract{The proper motions (PMs) of M31 and M33 are key to understanding the Local Group's dynamical evolution. However, measurement discrepancies between \textit{Gaia} blue and red samples, regarding whether the transverse velocity is remarkable, introduce significant ambiguity.
In this work, we remeasure the systemic PMs of M31 and M33 using massive supergiant stars from \textit{Gaia} Data Release 3. Clean disk tracers are selected via color--color diagrams, with foreground contaminants removed through kinematic and astrometric cuts.
We identify the discrepancy in M31’s blue and red samples as arising from systematic differences between \textit{Gaia}’s five-parameter (5p) and six-parameter (6p) astrometric solutions. The 6p solution, applied to sources lacking accurate color information, relies on a pseudo-color approximation, leading to lower precision and larger uncertainties. 
Two key limitations of the 6p solution are:  
1) degraded astrometric accuracy for very red sources ($G_{BP} - G_{RP} > 2.6$);  
2) significant PM zero-point offsets.
In our sample, red sources are dominated by the 6p solution, while blue sources include a substantial fraction of 5p sources; this mismatch drives the observed discrepancy.
By excluding extreme red sources and calibrating PM zero-points separately for 5p and 6p sources using background quasars, we reduce the discrepancy, bringing blue and red measurements into agreement within $1\sigma$. We ultimately report the most robust \textit{Gaia}-based PMs using high-quality 5p sources.
For M31, we obtain  
$(\mu_{\alpha}^{*}, \mu_{\delta})_{\rm M31} = (45.9 \pm 8.1,\ -20.5 \pm 6.6)\ \mu\mathrm{as}\,\mathrm{yr}^{-1}$,
consistent with, but more precise than, the HST result.  
For M33, we find that $(\mu_{\alpha}^{*}, \mu_{\delta})_{\rm M33} = (45.3 \pm 9.7,\ 26.3 \pm 7.3)\ \mu\mathrm{as}\,\mathrm{yr}^{-1}$, which agrees with the VLBA measurement within $1.5\sigma$. These results support a first infall scenario for M33.}

\keywords{Galaxies: Kinematics and Dynamics -- Local Group -- Proper Motion}
\maketitle
\nolinenumbers

\section{Introduction}\label{introduction}
The Milky Way (MW), Andromeda (M31), and Triangulum (M33) are the most massive members and dominant gravitational components of the Local Group (LG), whose total mass is estimated to be $(3$–$5)\times10^{12}\ M_{\odot}$ \citep[e.g.,][]{vdm2012a,Gonzales2014,Carlesi2022,Benisty2022}.  
The LG provides an ideal laboratory for investigating stellar formation and evolution, including studies of star formation histories \citep[e.g.,][]{Tolstoy2009,Weisz2014} and constraints on evolutionary tracks of massive stars \citep[e.g.,][]{Massey2003,Ekstrom2012}. It also serves as a crucial benchmark for models of hierarchical galaxy formation and cosmology \citep[e.g.,][]{McConnachie2012AJ,Klypin2002,Bullock2017}.  
Nevertheless, a key aspect remains uncertain: the relative motions of its three most massive galaxies.

Accurate proper motion (PM) measurements of M31, M33, and other LG satellites can significantly improve estimates of the LG's total mass and gravitational field, as well as the position and motion of its barycenter (e.g., \citealt{Kahn1959,Lynden-Bell1981,vdm2012a,Benisty2022}). This would allow for a more accurate positioning of the LG within the cosmic web and deepen our understanding of cosmological expansion, as the gravitational force competes with the cosmological expansion at LG's boundary.
On a smaller, galactic scale, precise measurements of the relative motion between M31 and the MW provide insights into their orbits and dynamical evolution, helping to distinguish between various formation histories, such as a first encounter or past accretion \citep{Hammer2013}, and predict their eventual merger (\citealt{vdm2012b,vdm2019,Schiavi2020,Sawala2024}). Similarly, accurate PM measurements of M31 and its potential satellite M33 are vital for understanding their orbital evolution, such as M33’s first infall scenario \citep{Patel2017}. These insights may also explain whether features such as the warp in the outer disk of M33 (\citealt{Corbelli2014,Kam2017}) and its tidal streams (\citealt{Bekki2008,McConnachie2009}) are caused by interactions with M31.

Although of vital importance, M31's PM has only been measured in recent decades due to its large distance.
The first direct measurement was made by \citet{Sohn2012} using the \textit{Hubble Space Telescope} (HST), based on multi-epoch observations of three deep fields (3.37\arcmin{} $\times$ 3.37\arcmin{}) over a baseline of 5--7 years. After accounting for internal motions, \citet{vdm2012a} derived a systemic PM of  
$(\mu_{\alpha}^{*}, \mu_{\delta})_{\rm M31} = (43.8 \pm 12.7,\ -31.7 \pm 12.2)\ \mu\mathrm{as}\,\mathrm{yr}^{-1}$, suggesting a nearly radial orbit between M31 and the MW. However, this estimate is strongly model-dependent \citep{vdm2012a}.  
Alternatively, M31’s PM can be inferred indirectly by modeling the line-of-sight (LOS) velocities of its satellite galaxies, under the assumption that the satellites are pressure-supported and share the bulk motion of M31 (e.g., \citealt{vdm2008, Salomon2016}). These methods, however, tend to favor a non-negligible transverse motion for M31, in contrast to the purely radial scenario.

Most recently, high-precision astrometric data from the {\it Gaia} mission have enabled direct measurements of M31’s PM using disk stars as tracers. \citet{vdm2019} demonstrated the feasibility of this approach by selecting nearly 1,000 disk stars from the color--magnitude diagram (CMD) constructed with {\it Gaia} Data Release 2 ({\it Gaia} DR2; \citealt{Gaiadr2}) broadband photometry. They derived a PM of  
$(\mu_{\alpha}^{*}, \mu_{\delta})_{\rm M31} = (65 \pm 32,\ -57 \pm 31)\ \mu\mathrm{as\,yr^{-1}}$, suggesting a significant transverse component comparable to the radial motion, albeit with large uncertainties due to the limited precision of {\it Gaia} DR2. Taking advantage of the substantially improved astrometric accuracy in the Early Data Release 3 ({\it Gaia} EDR3; \citealt{Gaiaedr3}), \citet{Salomon2021} refined this measurement. By incorporating \ion{H}{i} rotation models from \citet{Chemin2009} to correct for internal disk motions, they derived PMs using the blue disk sample and reported $(\mu_{\alpha}^{*}, \mu_{\delta})_{\rm M31} = (48.9 \pm 10.5,\ -36.9 \pm 8.1)\ \mu\mathrm{as\,yr^{-1}}$, in good agreement with the earlier HST result. However, measurements based on the red sample yielded significantly different values, implying a transversely dominated orbit for M31 relative to the MW.
Further analysis by \citet{Rusterucci2024}, who established a quasar-based reference frame and re-examined the same samples in four spatial regions, confirmed that the discrepancy between blue and red samples persists. This strongly suggests that the observed differences are not caused by spatial selection effects.

In the case of M33, the PM of its center of mass (COM) was first determined by \citet{Brunthaler2005} through Very Long Baseline Array (VLBA) observations of water masers. Thanks to its high angular resolution, the VLBA provided a robust measurement:  
$(\mu_{\alpha}^{*}, \mu_{\delta})_{\rm M33} = (23 \pm 7,\ 8 \pm 9)\ \mu\mathrm{as\,yr^{-1}}$.
Later, \citet{vdm2019} utilized data from {\it Gaia} Data Release 2 ({\it Gaia} DR2) and derived an estimate of $(\mu_{\alpha}^{*}, \mu_{\delta})_{\rm M33} = (31 \pm 35,\ -29 \pm 32)\ \mu\mathrm{as\,yr^{-1}}$. More recently, following a methodology similar to that used for M31, \citet{Rusterucci2024} updated the PM of M33 using {\it Gaia} DR3. Interestingly, the systematic discrepancy between blue and red stellar samples seen in M31 was not observed for M33.

Despite substantial progress over the past decade, the PMs of M31 and M33 remain uncertain. For M31, inconsistencies between different datasets, especially between the HST/{\it Gaia} blue samples and the {\it Gaia} red samples, have left the relative motion between the two most massive spirals in the LG ambiguous. To resolve this issue, further refined measurements based on purer stellar samples are necessary, as the bright sources selected from the {\it Gaia} color--magnitude diagram may be contaminated by foreground stars \citep{Salomon2021}.
 
In this paper, we resolve the discrepancy between {\it Gaia} blue and red samples and provide the most reliable {\it Gaia}-based PM measurements for M31 and M33. Our analysis utilizes pure supergiant samples, primarily selected from color-color or color--magnitude diagrams using $UBVRI$ or near-infrared (NIR) photometry, and eliminates contaminants through kinematic and astrometric criteria. The sample selection process is described in detail in Sect.~\ref{section:data}. By using supergiants with LOS velocity measurements, we establish the disk's rotation curve and derive the systemic PMs of M31 and M33, as outlined in Sect.~\ref{section:process}. The results, including an explanation for the origin of the discrepancy between {\it Gaia} blue and red samples, are presented in Sect.~\ref{section:result}. Finally, we summarize and discuss our findings in Sect.~\ref{section:summary_discussion}.

\section{Data}\label{section:data}
\subsection{Supergiant samples}\label{subsec:supergiants}
To provide accurate PM measurements, we selected reliable supergiant candidates as tracers.
In recent years, the number of supergiant candidates in M31 and M33 has significantly increased. This is partly due to the effectiveness of using the Johnson Q-index to identify blue supergiants and NIR CMDs to identify red supergiants, which successfully distinguish supergiants from foreground dwarfs. More importantly, LOS velocities obtained from wide-field multi-object fiber spectrographs and high-precision astrometric measurements from {\it Gaia} provide two independent methods for robustly excluding foreground contaminants, especially for yellow supergiants that heavily overlap with foreground stars in the CMD (\citealt{Drout2009,Drout2012,Massey2016rsg,Massey2016,Massey2021,Wu2024}). 

In this work, our sample of supergiants was compiled from four sources:
\begin{enumerate}
    
    \item \citet{Massey2016} catalog: This compilation includes previous identifications by Massey’s group and incorporates follow-up spectroscopic observations with the Hectospec spectrograph on the MMT. We selected objects labeled as \texttt{members}, \texttt{possible members}, or \texttt{unknown}, excluding those identified as H\,\textsc{ii} regions or clusters.
    
    \item Red supergiants from \citet{Massey2021}: Red supergiant candidates identified via NIR CMDs, with foreground contaminants excluded using \textit{Gaia} DR2 astrometry.

    \item \citet{Wu2024} catalog: This catalog provides systematic identifications of supergiants in M31 and M33. They selected candidates based on photometric criteria, then removed contaminants using LOS velocities from the Large Sky Area Multi-Object Fiber Spectroscopic Telescope (LAMOST; \citealt{Cui2012}) and astrometric data from \textit{Gaia} DR3. We retained objects classified as \texttt{Rank1} and \texttt{Rank2}.
    
    \item Blue supergiants from LGGS photometry: We supplemented the blue end using photometric data from the Local Group Galaxies Survey (LGGS; \citealt{Massey2006,Massey2016}). We selected sources with Johnson $Q$-index $<$ $-0.6$, a reddening-free indicator of intrinsic color. Previous studies have shown this selection yields a clean sample with minimal foreground contamination \citep{Massey2006}. To eliminate non-stellar objects, we crossmatched with catalogs including \citet{Galleti2004,Galleti2007}, \citet{Merrett2006}, \citet{Azimlu2011}, \citet{Sanders2012}, and \citet{Zhang2020} for M31, and \citet{Sarajedini2007}, \citet{Ciardullo2004}, \citet{Hodge1999}, and \citet{Zhang2020} for M33. Finally, we retained sources with low reddening, defined as $G_{BP} - G_{RP} < 0.5$.
\end{enumerate}

After crossmatching with \textit{Gaia} DR3 within a radius of 2\arcsec, we obtained a total of 10,173 supergiant candidates in M31 and 10,883 in M33. Among them, 8,450 candidates in M31 and 8,691 in M33 have available PM measurements (i.e., a five-parameter or six-parameter astrometric solution).

\subsection{Sample selection}\label{subsec:cleaning}
To improve the purity of the disk tracer sample, we applied a cleaning procedure to the supergiant candidates. This involved a series of selection criteria designed to remove potential contaminants and sources with unreliable astrometric measurements. The specific steps of this procedure, along with the number of remaining sources after each step, are summarized in Tables~\ref{tab:m31_supergiants} and~\ref{tab:m33_supergiants}.

\begin{table}[htbp!]
\centering
\caption{Summary of the cuts applied to clean up the M31 supergiant samples and the corresponding number of sources remaining.}
\label{tab:m31_supergiants}
\begin{tabular}{ll}
\hline
Cuts & Remaining sources \\
\hline
Total & 8,450 \\
1) 1.8$^\circ$ Projected ellipse & 8,397 \\
2) Parallax Cut & 7,615 \\
3) Non-binary & 6,614 \\
4) Flux excess Restriction & 4,829 \\
5) G $<$ 20 & 2,880 \\
6) PM Cuts & 2,462 \\
\hline
\end{tabular}
\end{table}

\begin{table}[htbp!]
\centering
\caption{Similar to Table~\ref{tab:m31_supergiants}, but for M33.}
\label{tab:m33_supergiants}
\begin{tabular}{ll}
\hline
Cuts & Remaining sources \\
\hline
Total & 8,691 \\
1) 1.0$^\circ$ Projected ellipse & 8,691 \\
2) Parallax Cut & 8,076 \\
3) Non-binary & 5,674 \\
4) Flux excess Restriction & 4,039 \\
5) G $<$ 20 & 2,551 \\
6) PM Cuts & 2,282 \\
\hline
\end{tabular}
\end{table}

For M31, we adopted the following basic parameters: COM position (RA, Dec) = (10.68$^\circ$, 41.27$^\circ$) \citep{Evans2010}, inclination $i = 77.5^\circ$ (\citealt{Walterbos1988,deVaucouleurs1991}), and position angle ${\rm PA} = 37.5^\circ$ \citep{Chemin2009,Corbelli2010}. We then applied a series of selection criteria to exclude contaminants and sources with poor measurements, as is detailed below:

Step 1. Taking the angular size of M31's disk to be 1.8$^\circ$, its projected shape on the sky is approximated by an ellipse. We excluded all sources lying outside this elliptical boundary, resulting in a sample of 8,397 sources for further cleaning.

Step 2. We corrected the parallax zero-point following \citet{Lindegren2021par}, and removed sources whose distances exceed that of M31 (784 kpc; \citealt{Stanek1998}) at the 2$\sigma$ confidence level. Specifically, we retained sources satisfying $\varpi - 2\sigma_\varpi < 1/784$, leaving 7,615 sources.

Step 3. To exclude potential non-single stars, we applied three astrometric quality criteria from \citet{Fabricius2021}:  
\begin{enumerate}
    \item ${\rm RUWE} < 1.4$; \notag
    \item \texttt{ipd\_frac\_multi\_peak} $<$ 2; \notag
    \item \texttt{ipd\_gof\_harmonic\_amplitude} $<$ 0.1. \notag
\end{enumerate}
This step reduced the sample to 6,614 sources.

Step 4. To mitigate the effects of flux excess in crowded or faint regions, we adopted the criterion from \citet{Riello2021}, and retained sources whose corrected excess factor ($C_{\rm corr}$) satisfies
\begin{equation}
    \label{Eq:corr}
    |C_{\rm corr}| < 3 \times \left(5.9898 \times 10^{-3} \times 8.817481 \times 10^{-12} \times G^{7.618399} \right).
\end{equation}
After this cut, 4,829 sources remained.

Step 5. We excluded sources with $G$-band magnitudes fainter than 20 mag to ensure the reliability of astrometric measurements, yielding 2,880 sources.

Step 6. Finally, we applied PM cuts:  
$|\mu_{\alpha}^{*}| < 0.19\,{\rm mas\,yr^{-1}} + 2\sigma_{\mu_{\alpha}^{*}}$ and  
$|\mu_\delta| < 0.19\,{\rm mas\,yr^{-1}} + 2\sigma_{\mu_\delta}$,  
where $\sigma_{\mu_{\alpha}^{*}}$ and $\sigma_{\mu_\delta}$ are the PM uncertainties in right ascension and declination, respectively. The threshold of 0.19 ${\rm mas\,yr^{-1}}$ follows the criterion adopted in \cite{Salomon2021} and \cite{Rusterucci2024}. These limits correspond to rejecting sources with transverse velocities exceeding around 750\,km\,s$^{-1}$ at the distance of M31 in either direction. This final selection yielded 2,462 sources, which are considered a clean sample of disk members for subsequent analysis.

For M33, we applied a screening process similar to that used for M31, with appropriate adjustments. The adopted basic parameters were: COM position (RA, Dec) = (23.46$^\circ$, 30.66$^\circ$) \citep{van2000}, inclination $i = 54^\circ$ \citep{deVaucouleurs1991}, and position angle ${\rm PA} = 23^\circ$ (\citealt{deVaucouleurs1991,McConnachie2006}). Contaminants and sources with poor-quality measurements were excluded through the following steps:

Step 1. Assuming an angular size of 1.0$^\circ$ for M33's disk, its projected shape on the sky is approximated by an ellipse, within which all 8,691 sources were located.

Step 2. Adopting a distance to M33 of 840 kpc \citep{Galleti2004distance,Breuval2023}, we retained sources that satisfy $\varpi - 2\sigma_\varpi < 1/840$, yielding 8,076 sources.

Steps 3--5. The same astrometric quality and photometric excess criteria as applied for M31 were used here. The number of sources remaining after each step is listed in Table~\ref{tab:m33_supergiants}.

Step 6. The PM limits were adjusted to  
$|\mu_{\alpha}^{*}| < 0.14\,{\rm mas\,yr^{-1}} + 2\sigma_{\mu_{\alpha}^{*}}$ and  
$|\mu_\delta| < 0.14\,{\rm mas\,yr^{-1}} + 2\sigma_{\mu_\delta}$,  
corresponding to a transverse velocity threshold of about 550\,km\,s$^{-1}$ at the distance of M33. 
This threshold is similar to that adopted in \cite{Salomon2021} and \cite{Rusterucci2024}, although it is slightly more stringent.
We consider it more robust given M33’s lower disk rotation velocity (see Sect.~\ref{subsect:rc}) and shallower gravitational potential compared to M31.
The final cut left 2,282 sources, which are considered a clean sample of disk members for subsequent analysis.

\section{Analysis}\label{section:process}
\subsection{Rotation curve from disk objects}\label{subsect:rc}
To accurately assess the contribution of disk rotation to the observed PM of supergiant stars, we constructed rotation curve (RC) models for the disks of M31 and M33. The modeling samples were selected from those identified through the six steps described above (see Tables~\ref{tab:m31_supergiants} and~\ref{tab:m33_supergiants}), with the additional requirement that only stars with available heliocentric LOS velocity measurements were included. 
This selection ensures that the samples used to construct the RCs are both clean and robust.

\subsubsection{M31 rotation curve}\label{subsec:M31RC}

\begin{figure*}[htbp!]
\centering
    \begin{minipage}{0.49\textwidth}
    \centering
    \includegraphics[width=\textwidth]{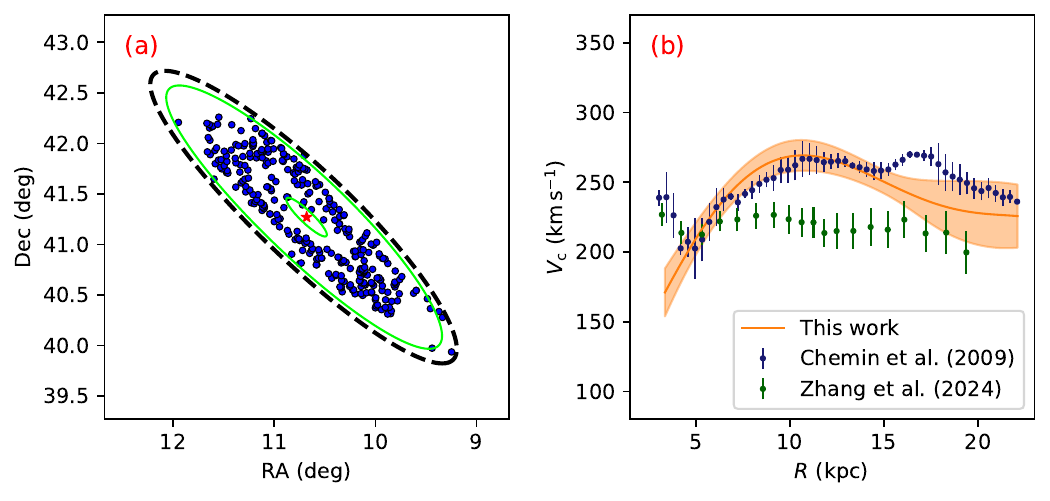}
    \end{minipage}
    \hspace{0.01\textwidth}
    \begin{minipage}{0.49\textwidth}
    \centering
    \includegraphics[width=\textwidth]{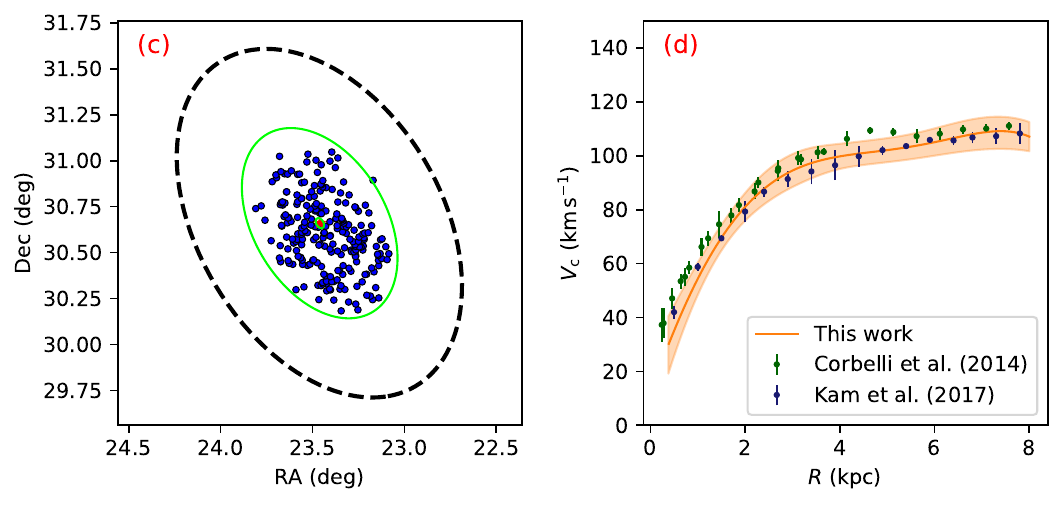}
    \end{minipage}
\caption{Panel (a): Spatial distribution of 321 supergiant candidates for modeling the RC of M31 disk region. The red star denotes the COM of M31, and the dashed black outer line represents the 1.8$^\circ$ ellipse of the projected disk. Inner and outer green ellipses denote the reliable range of our derived RC. Panel (b): RC of the M31 disk region. The orange line and the shaded area represent the best fit and the corresponding standard error from this work, while the blue and green error-bars represent results from previous studies, as labeled in the bottomright corner.
Panel (c): Spatial distribution of 235 supergiant candidates for modeling the RC of M33 disk region. The red star denotes the COM of M33, and the dashed black outer line represents the 1$^\circ$ ellipse of the projected disk. Inner and outer green ellipses denote the reliable range of our derived RC. Panel (d): RC of the disk of M33. The orange line and the shaded area illustrate the best fit and its corresponding standard error from this work. The blue and green error-bars represent results obtained from previous studies by \ion{H}{i} observations, as labeled in the bottomright corner.}
\label{fig:RC_m31_m33}
\end{figure*}

For M31, among the 2,462 sources listed in Table~\ref{tab:m31_supergiants} after applying the six selection steps, 321 have LOS velocity measurements from \cite{Drout2009}, \cite{Massey2009}, \cite{Cordiner2011ApJ}, \cite{Massey2016rsg}, and \cite{Wu2024}. As young tracers, these supergiant candidates experience negligible asymmetric drift, which can be ignored in the RC calculations. Since the LOS velocities from different catalogs were obtained using different instruments, we applied the zero-point offsets derived by \cite{Wu2024} to adjust the MMT LOS velocities to the LAMOST scale. The spatial distribution of these 321 sources is shown in Panel (a) of Fig.~\ref{fig:RC_m31_m33}.

Equipped with observed LOS velocities, we followed a procedure similar to \cite{Rubin1970} to derive the RC. 
The LOS velocity of a disk target can be expressed as
\begin{equation}
    \label{Eq:RC_M31}
     V_{\rm LOS} = V_{\rm LOS}^{\rm M31} + V_{c}(R){\sin{i}}{\cos{\theta}}\text{.}
\end{equation}
Here, $\cos{\theta}=X/R$, where $X$ represents the distance along the semimajor axis and $R$ is the radial distance of the object within the M31 disk plane, and $i$ is the inclination angle setting to be $77.5^{\circ}$ \citep[]{Walterbos1988,deVaucouleurs1991}. $V_{\rm LOS}^{\rm M31}$ is the systemic LOS velocity of M31 and $V_{c}$ is the circular speed at $R$.

Instead of assuming a constant $V_c$, we modeled the RC using a fifth-order polynomial to account for potential variations, following the approach adopted in previous studies \citep[e.g.,][]{Rubin1970,Rastorguev2017}.
A Markov chain Monte Carlo (MCMC) method was employed to simultaneously fit both $V_{\rm LOS}^{\rm M31}$ and the coefficients of the polynomial.
To facilitate the fitting, the sources were divided into eight radial bins, each containing at least 10 stars.
We derived a $V_{\rm LOS}^{\rm M31}$ of $-300.8 \pm 1.2$ km\,s$^{-1}$, consistent with previous measurements (\citealt{vdm2008}, \citealt{Watkins2013}, \citealt{Salomon2016}).

For the RC construction, we consider results outside the radial range of 3.35 to 22.1 kpc from the M31 center to be unreliable due to the limited number of sources. 
These boundaries are indicated as green ellipses in Panel (a) of Fig.~\ref{fig:RC_m31_m33}.
The RC within this range, together with previous measurements from \cite{Chemin2009} and \cite{Zhang2024}, is shown in Panel (b) of Fig.~\ref{fig:RC_m31_m33}.
Overall, our RC is consistent with previous results within the 3$\sigma$ confidence level.
Specifically, it closely follows the \ion{H}{i} observations of \cite{Chemin2009} between around 4 and 15 kpc and then trends toward an intermediate position between their results and the latest measurements from \cite{Zhang2024}, which are based on multiple tracers.
The final expression for our RC is
\begin{align}
    V_c(R) &= -0.0004109 \times R^{5} + 0.02668 \times R^{4} \notag \\
    &- 0.5688 \times R^{3} + 3.470 \times R^{2} + 17.23 \times R + 92.44\text{.} \notag
\end{align}

With the RC in hand, we can precisely determine the contribution of disk rotation to the motion of each disk object. Accordingly, we revised the selection criteria from Step 6 in Sect.~\ref{subsec:cleaning} as
\begin{align}
    &|\mu_{\alpha}^{*} - \mu_{\alpha,\rm RC}^{*}| < 0.14\,{\rm mas\,yr^{-1}} + \sigma_{\mu_{\alpha}^{*}} \notag, \\
    &|\mu_{\delta} - \mu_{\delta, \rm RC}| < 0.14\,{\rm mas\,yr^{-1}} + \sigma_{\mu_{\delta}} \notag,
\end{align}
where $\mu_{\alpha, \rm RC}^{*}$ and $\mu_{\delta, \rm RC}$ are the predicted PM components arising solely from disk rotation, as derived from our RC models. Given the significant internal kinematic variations across M31 and their contributions to the observed PMs, these revised criteria are more physically motivated than those in Step~6 originally proposed by \citet{vdm2019} and \citet{Salomon2021}, and effectively remove nonphysical outliers and hyper-velocity stars with deviations exceeding about 500\,${\rm km\,s^{-1}}$ from the expected RC. We note that a $1\sigma$ threshold was adopted for the cuts, instead of the $2\sigma$ used in the original Step~6, to more effectively exclude contaminants or sources with large uncertainties. 

Using above cuts and selecting only supergiant stars within the radius range of 3.35 to 22.1 kpc (the valid range of the RC model), a total of 1,587 sources remain. Among them, 783 are blue sample stars (blue and yellow supergiants) and 804 are red sample stars (red supergiants). The spatial distribution and CMD positions of these sources are shown in the left-hand panels of Fig.~\ref{fig:ini_m31_m33_distribution}.

\subsubsection{M33 rotation curve}\label{subsec:M33RC}
For M33, 235 out of the 2,282 sources remaining after the six selection steps (Table~\ref{tab:m33_supergiants}) have heliocentric LOS velocity measurements \citep{Drout2012, Wu2024}. Again, we applied a zero-point correction to the MMT LOS velocities to align their scale with that of LAMOST. The spatial distribution of these sources is shown in Panel (c) of Fig.~\ref{fig:RC_m31_m33}.

Following the procedure similar to that used for M31, we derived the RC of M33’s disk region using the LOS velocities of disk objects, which can be predicted from RC by
\begin{equation}
    \label{Eq:RC_M33}
     V_{\rm LOS} = V_{\rm LOS}^{\rm M33} + V_{c}(R){\sin{i}}{\cos{\theta}}\text{.}
\end{equation}
Here the inclination angle $i$ was adopted as $54.0^{\circ}$ from \citet{deVaucouleurs1991}.
The definitions of $V_{\rm LOS}^{\rm M33}$ and $\theta$ are similar to those in Eq.~\ref{Eq:RC_M31} for M31.
Again, the RC was modeled using a fifth-order polynomial.
The best fitting results were then derived using an MCMC method. The derived $V_{\rm LOS}^{\rm M33}$ is $-$179.5$\pm$0.6\,km\,s$^{-1}$, in line with previous studies (\citealt{vdm2008,Kam2017}). The RC is considered reliable within 0.38–8 kpc from M33’s center, as indicated by the green ellipses in Panel (c) of Fig.~\ref{fig:RC_m31_m33}. Within this range, our results agree with previous \ion{H}{i}-based RC measurements by \citet{Corbelli2014} and \citet{Kam2017}, particularly showing strong consistency between 1 and 8 kpc (Panel (d) of Fig.~\ref{fig:RC_m31_m33}).
The best-fit RC is expressed as
\begin{align}
    V_c(R) &= -0.01626\times R^{5}+0.2555\times R^{4} \notag \\
    &-0.6677\times R^{3}-7.518\times R^{2}+50.59\times R+11.7\notag.
\end{align}

\begin{figure*}[htbp]
\centering
    \begin{minipage}{0.49\textwidth}
    \centering
    \includegraphics[width=\textwidth]{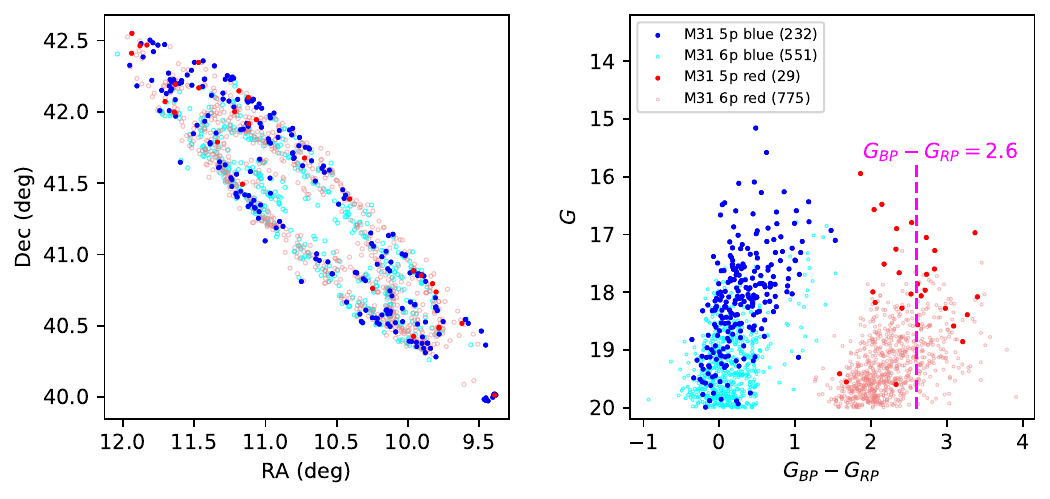}
    \end{minipage}
    \hspace{0.01\textwidth}
    \begin{minipage}{0.49\textwidth}
    \centering
    \includegraphics[width=\textwidth]{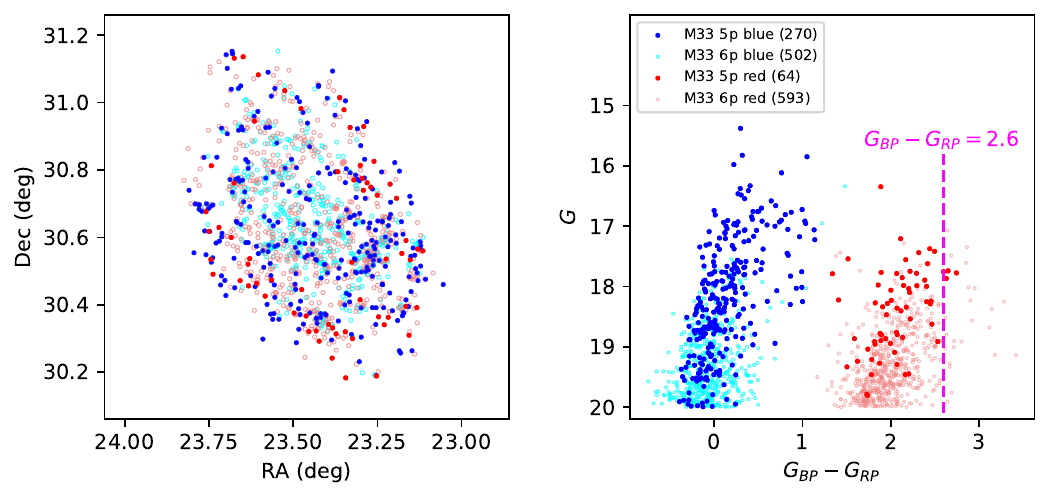}
    \end{minipage}
\caption{Spatial and CMD distributions of final M31 and M33 supergiant samples (see Sect.~\ref{subsect:rc}). The \textit{left-hand panels} display the spatial and CMD distributions of the M31 sample stars, while the \textit{right-hand panels} present those of the M33 sample stars. Filled blue and red symbols denote blue (including some yellow) and red supergiants with a 5p astrometric solution, respectively; open cyan and pink circles indicate blue and red supergiants with a 6p solution. The dashed magenta line in the CMD panels marks \( G_{BP} - G_{RP} = 2.6 \). Clearly, M31 contains a significantly larger population of extremely red supergiant stars compared to M33.}
\label{fig:ini_m31_m33_distribution}
\end{figure*}

For the same reason as for M31, we revised the criteria in Step 6 of Sect.~\ref{subsec:cleaning} for M33 as
\begin{align}
    &|\mu_{\alpha}^{*} - \mu_{\alpha,\rm RC}^{*}| < 0.11\,{\rm mas\,yr^{-1}} + \sigma_{\mu_{\alpha}^{*}}\notag, 
    \\
    &|\mu_{\delta} - \mu_{\delta, \rm RC}| < 0.11\,{\rm mas\,yr^{-1}} + \sigma_{\mu_{\delta}} \notag.
\end{align}
After applying these revised cuts, a total of 1,429 sources remain within the radial range of 0.38 to 8 kpc (the valid range for the RC model). This includes 772 blue sample stars (blue and yellow supergiants) and 657 red sample stars (red supergiants). Their spatial distribution and CMD positions are shown as blue and red dots in the right-hand panels of Fig.~\ref{fig:ini_m31_m33_distribution}.

\subsection{Method for proper motion determination}\label{subsec:mcmc}
The method adopted here to determine the systemic PMs of M31 and M33 is generally similar to those used in previous studies, such as \citet{Salomon2021} and \citet{Rusterucci2024}.
The dispersion and asymmetric drift are neglected in our modeling given the young, kinematically cold nature of the massive supergiant stars used in this study. In other words, the motions of these supergiants are assumed to follow perfectly circular orbits within the disks of M31 and M33.
Under this assumption, the expected PM of a supergiant consists of two components: one arising from its circular motion within the disk, and the other representing the systemic PM of the galaxy. This can be written as
\begin{equation}
\label{Eq:model}
\boldsymbol{\mu}^{\rm model} = \boldsymbol{\mu}_{\text{disk}} + \boldsymbol{\mu}_{\text{sys}},
\end{equation}
where \(\boldsymbol{\mu}_{\text{disk}}\) is the contribution from the circular motion, which can be precisely predicted at any position within M31 and M33 based on the RCs derived in Sects.~\ref{subsec:M31RC} and~\ref{subsec:M33RC}.  
\(\boldsymbol{\mu}_{\text{sys}}\) denotes the systemic PMs of M31 and M33, respectively, which are the quantities we aim to determine.
To infer the systemic PMs of M31 and M33, we used an MCMC method to sample the posterior distribution.
The probability density function for one tracer at position $(\alpha_k, \delta_k)$ can be written as
\begin{align}
    \label{Eq:pdf}
    f(\mu_{\alpha,k}^{*},\mu_{\delta,k}) &= \frac{1}{2\pi\sigma_{\mu_{\alpha,k}^{*}}\sigma_{\mu_{\delta,k}}\sqrt{1-\rho_{k}^2}} \times \notag \\
    &\exp\left(-\frac{1}{2(1-\rho_{k}^2)} \left[\frac{\Delta_{\mu_{\alpha,k}^{*}}^2}{\sigma_{\mu_{\alpha,k}^{*}}^2} + \frac{\Delta_{\mu_{\delta,k}}^2}{\sigma_{\mu_{\delta,k}}^2} - \frac{2\rho_{k}\Delta_{\mu_{\alpha,k}^{*}}\Delta_{\mu_{\delta,k}}}{\sigma_{\mu_{\alpha,k}^{*}}\sigma_{\mu_{\delta,k}}}\right]\right)\text{,}
\end{align}
where $\Delta_{\mu_{\alpha,k}^{*}}$ and $\Delta_{\mu_{\delta,k}}$ are the differences between the observed PM of the \(k\)-th supergiant star and the modeled value given by Eq.~\ref{Eq:model}.
Here, \(\sigma_{\mu_{\alpha,k}^{*}}\) and \(\sigma_{\mu_{\delta,k}}\) are the PM uncertainties in right ascension and declination, respectively, taken from {\it Gaia} DR3.  
In addition, the measured correlation between the two components, \(\rho_k \equiv \texttt{pmra\_pmdec\_corr}\), is also taken into account.
To evaluate the impact of potential residual contamination on the derived PMs, we first adopted a likelihood function that includes a contamination component, following the functional form used in \citet{Salomon2021}. In this model, the observed PM distribution is treated as a mixture of two populations, true M31 members and contaminants, with the likelihood given by  
\begin{equation}
\label{Eq:likelihood_cont}
\ln \mathcal{L} = \sum_{k=1}^{n} \ln \left[ (1-\eta_{c})f(\mu_{\alpha,k}^{*,t}, \mu_{\delta,k}^{t}) + \eta_{c}g(\mu_{\alpha,k}^{*,c}, \mu_{\delta,k}^{c}) \right],
\end{equation}
where \(\eta_{c}\) is the contamination fraction, \((\mu_{\alpha,k}^{*,t}, \mu_{\delta,k}^{t})\) represent the PMs of true members, and \((\mu_{\alpha,k}^{*,c}, \mu_{\delta,k}^{c})\) represent those of contaminants. The contaminant probability density function, \(g\), has the same form as Eq.~\ref{Eq:pdf}, but with distinct mean and dispersion parameters from those of the true members. Following \citet{Salomon2021}, we imposed uniform priors on the contaminant dispersions: \(0 \leq \sigma_{\mu_{\alpha}}^{*,c},\, \sigma_{\mu_{\delta}}^{c} \leq 2\,{\rm mas\,yr^{-1}}\).  
We performed MCMC sampling using this likelihood and found that the inferred contamination fraction is below 0.1\%, indicating that our supergiant sample is highly pure and any residual contaminants have negligible impact on the PM results. Therefore, for the subsequent analysis, we adopted a simplified likelihood function without the contamination component to infer the systemic PM via the MCMC technique:  
\begin{equation}
\label{Eq:likelihood}
\ln \mathcal{L} = \sum_{k=1}^{n} \ln f(\mu_{\alpha,k}^{*}, \mu_{\delta,k}).
\end{equation}

\section{Result and discussion}\label{section:result}
\subsection{Origin of the discrepancy between M31 blue and red samples}\label{subsec:origin of discrepancy}

\begin{figure*}[ht!]
\centering
    \begin{minipage}{0.49\textwidth}
    \centering
    \includegraphics[width=\textwidth]{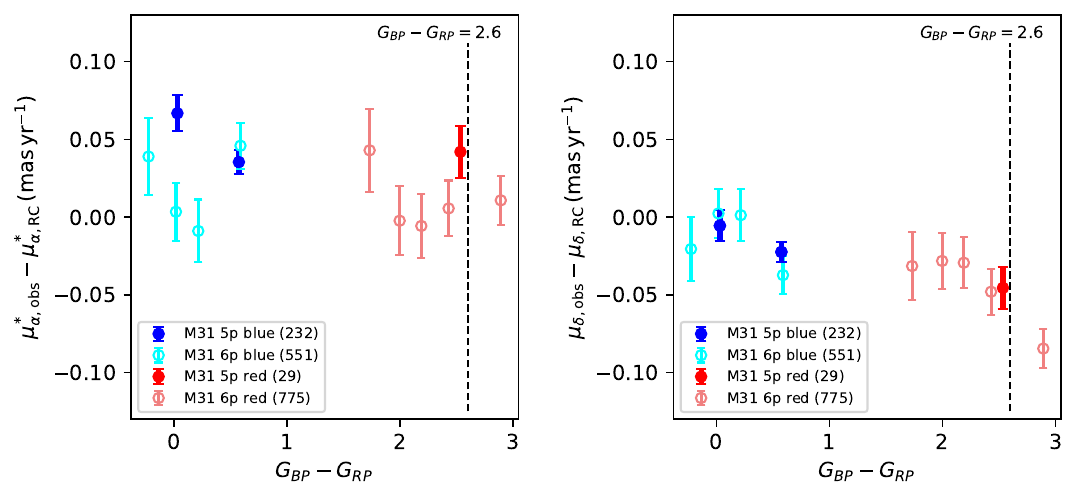}
    \end{minipage}
    \hspace{0.01\textwidth}
    \begin{minipage}{0.49\textwidth}
    \centering
    \includegraphics[width=\textwidth]{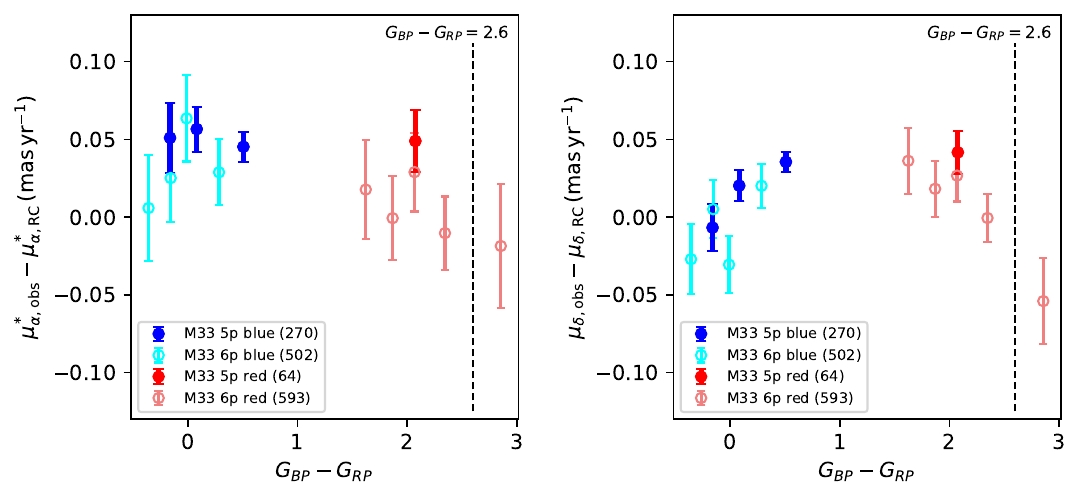}
    \end{minipage}
\caption{\textit{Left-hand panels}: PM distribution of the blue and red M31 samples as a function of $G_{BP} - G_{RP}$. The data points in the left (right) panel show the weighted average differences between the observed PMs, $\mu_{\alpha, \mathrm{obs}}^{*}$ ($\mu_{\delta, \mathrm{obs}}$), from {\it Gaia} DR3 and the predicted rotational components $\mu_{\alpha, \rm RC}^{*}$ ($\mu_{\delta, \rm RC}$), derived from the RC in Sect.~\ref{subsec:M31RC}. Weights are proportional to $1/\sigma_{\alpha*}^2$ or $1/\sigma_{\delta}^2$, and the errors, also weighted by uncertainty, are indicated accordingly.
\textit{Right-hand panels}: Similar to the \textit{left-hand panels}, but for M33. }
\label{fig:pm_bprp_m31}
\end{figure*}

Using the carefully selected supergiant star samples and the established methodology described above, we remeasured the PM of M31. As is shown clearly in Fig.~\ref{fig:ini_m31_m33_distribution}, the supergiant star sample comprises two distinct populations: blue and red supergiant stars\footnote{Yellow supergiant stars are largely absent, likely due to two factors: 1) the intrinsically short duration of this evolutionary phase for massive stars, and 2) low identification efficiency resulting from heavy foreground contamination.}.
These populations are broadly consistent with those used in previous studies \citep{Salomon2021, Rusterucci2024}.
We first derived the PM of M31 using the two subsamples separately. From the blue supergiants, we obtained $(\mu_{\alpha}^{*}, \mu_{\delta})_{\rm M31,Blue} = (37.7 \pm 5.2, -17.0 \pm 4.4)\,\mu{\rm as\,yr^{-1}}$, while the red supergiants yielded $(\mu_{\alpha}^{*}, \mu_{\delta})_{\rm M31,Red} = (15.3 \pm 7.7, -50.3 \pm 6.3)\,\mu{\rm as\,yr^{-1}}$. The PM derived from the red sample is significantly large, indicating a non-negligible transverse motion of M31, whereas the result from the blue sample suggests a nearly negligible tangential component.
This apparent discrepancy between the two results echoes similar tensions reported by \citet{Salomon2021} and \citet{Rusterucci2024}.

As for the cause of this discrepancy, \citet{Rusterucci2024} investigated the influence of spatially dependent zero-point offsets using quasars (QSOs) and found that the mean motion remains close to 0\,$\mu{\rm as\,yr^{-1}}$. Further analysis, dividing the sky into four quadrants, shows that the discrepancy between the blue and red supergiant samples persists across all regions, suggesting that spatial systematics are unlikely to be the dominant cause.

Instead, we recall two key aspects highlighted by the {\it Gaia} team during astrometric validation. The first concerns the nature of the astrometric solutions in {\it Gaia} (E)DR3, which include two-parameter (position only), five-parameter (5p, position, parallax, and PM), and six-parameter (6p, 5p plus pseudo-color) solutions \citep{Lindegren2021ast}. The 6p solution is typically adopted when the effective wavenumber ($\nu_{\rm eff}$), which provides essential color information for correcting chromatic effects in the point-spread function, cannot be reliably determined, particularly in crowded fields or areas affected by bright stars. In such cases, a pseudo-color ($\hat{\nu}_{\rm eff}$) is introduced and solved simultaneously with the five standard astrometric parameters. As a result, the 6p solution generally yields less precise astrometry and exhibit both larger random and systematic uncertainties than the 5p solution, due to more complex observational conditions and additional correlations introduced by the pseudo-color term \citep{Lindegren2021ast, Fabricius2021}.
The second factor is the color dependence of astrometric quality. Based on open cluster members as calibrators, \citet{Fabricius2021} demonstrated that redder faint sources ($G > 18$) show significant systematic offsets in PM relative to the cluster median, particularly toward the red end of the color distribution ($G_{BP}-G_{RP} > 2.0$; see their Figs.~17, 18, 24, and 25).

Given the two potential issues discussed above, we began by examining the composition of our samples in terms of 5p and 6p astrometric solutions. In M31, the blue supergiant sample contains 232 stars with a 5p solution and 551 with a 6p one, whereas the red supergiant sample is heavily dominated by 6p sources, with 775 stars having a 6p solution and only 29 having a 5p one. A similar pattern is observed in M33, where the red sample again shows a significantly higher proportion of 6p sources compared to the blue. The spatial and CMD distributions of these stars, categorized by both color and astrometric solution type, are presented in Fig.~\ref{fig:ini_m31_m33_distribution}.

To assess whether the observed discrepancies are linked to systematics associated with astrometric solution type or stellar color, we examined the PM distribution as a function of $G_{BP}-G_{RP}$ for four subsamples in M31: blue 5p, blue 6p, red 5p, and red 6p, as is shown in the left-hand panels of Fig.~\ref{fig:pm_bprp_m31}. In this analysis, the PM values have been corrected for disk rotation. For the blue and red 6p subsamples, each color bin contains at least 130 and 150 sources, respectively, spanning the $G_{BP}-G_{RP}$ range in steps of more than 0.2 mag. The blue 5p subsample is similarly binned, with each comprising at least 115 sources. The filled red circle represents the 29 red stars with a 5p solution, while the last open pink circle denotes 155 red 6p sources with $G_{BP}-G_{RP} > 2.6$.
Two notable trends can be found:
\begin{enumerate}
    \item Among the 6p sources, PM values exhibit a marked discontinuity at the red extreme ($G_{BP}-G_{RP} > 2.6$), with significant deviations from the rest of the sample, particularly in the $\delta$ direction. The spatial distribution of these sources is relatively uniform, suggesting that the offset is unlikely to originate from spatial effects. Instead, this likely reflects systematic errors in the {\it Gaia} 6p astrometric solution. This finding is consistent with \citet{Fabricius2021}, who reported substantial PM offsets in faint, red sources, populations typically dominated by the 6p solution given their faint nature (most with $G > 18$; \citealt{Lindegren2021ast}). Due to the limited number of red stars with a 5p solution, it remains unclear whether they are similarly affected. However, the 29 red 5p stars, despite having a mean color near $G_{BP}-G_{RP} = 2.6$, exhibit PM values comparable to those of bluer stars with either a 5p or 6p solution. This suggests that the 5p solution, regardless of color, is generally reliable in our analysis.
    \item A systematic offset is also observed between 5p and 6p solutions: stars with a 5p solution exhibit systematically higher PM values in the $\alpha$ direction compared to their 6p counterparts, independent of color.
\end{enumerate}

We then examined the PM distribution for M33 using the same approach. As is shown in the right-hand panels of Fig.~\ref{fig:pm_bprp_m31}, the blue sample includes at least 90 (125) stars per bin for the 5p (6p) subsamples, with $G_{BP}-G_{RP}$ binned over intervals of 0.2 (0.125) mag. For the red 6p sample with $G_{BP}-G_{RP} \leq 2.6$, each bin contains at least 125 stars in 0.2 mag intervals. Additionally, the final open pink circle represents the 23 red 6p stars with $G_{BP}-G_{RP} > 2.6$. The overall patterns observed in M33 are consistent with those found in M31.

From these tests, we conclude that the discrepancy between the blue and red samples primarily arises from systematic issues associated with the 6p astrometric solution. These issues are twofold: (1) unreliable astrometry for sources at the extreme red end, particularly those with $G_{BP}-G_{RP} > 2.6$, and (2) systematic offsets between 5p and 6p solutions, regardless of color. These effects are further amplified by the differing proportions of 5p and 6p solutions in the two samples: the red sample is dominated by 6p sources, while the blue sample includes a substantial fraction of 5p sources (approximately 30\% in M31 and 35\% in M33). As a result, the red sample is more vulnerable to the systematics of the 6p solution.

To address these two issues, we first mitigated the impact of unreliable astrometry by restricting the 6p sample to sources with $G_{BP} - G_{RP} \leq 2.6$. Second, to account for systematic offsets, we examined potential differences in the zero-point corrections between the 5p and 6p solutions using QSOs, as described below.

\subsection{Proper motion zero-point offset}\label{subsec:zero-point}

\begin{figure*}[htbp!]
\centering
    \begin{minipage}{0.49\textwidth}
    \centering
    \includegraphics[width=\textwidth]{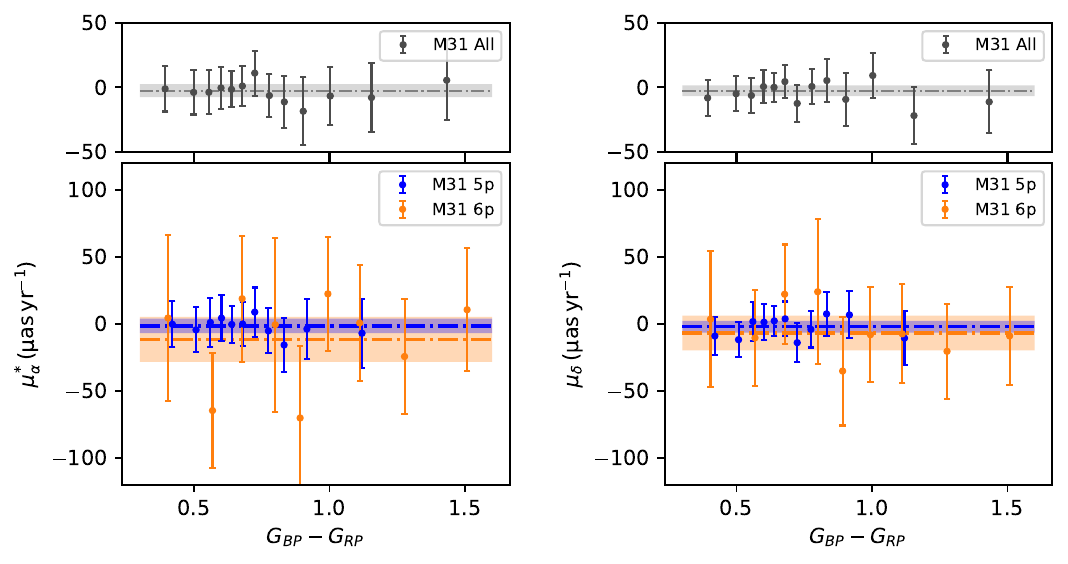}
    \end{minipage}
    \hspace{0.01\textwidth}
    \begin{minipage}{0.49\textwidth}
    \centering
    \includegraphics[width=\textwidth]{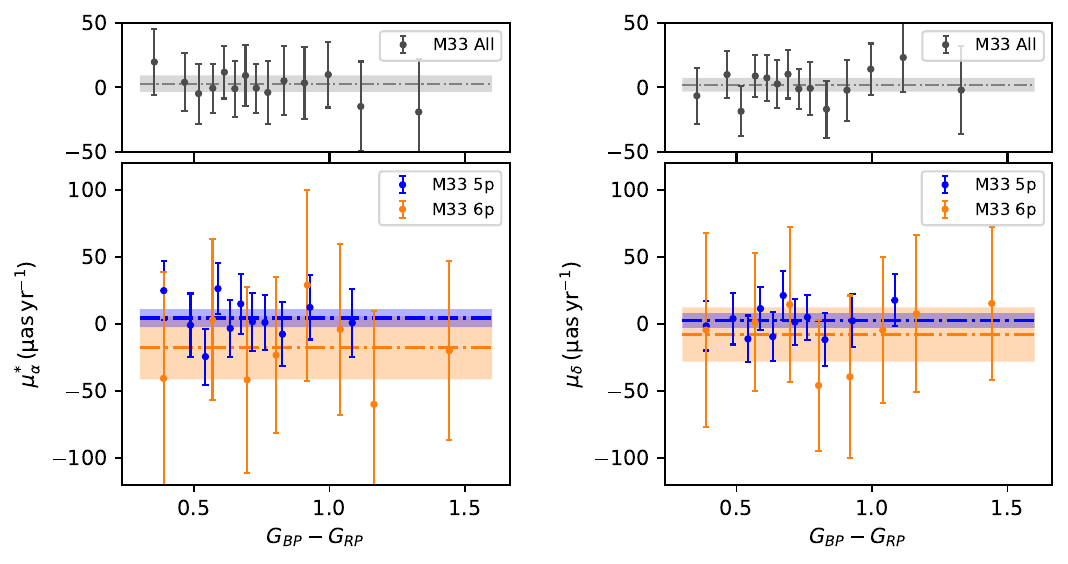}
    \end{minipage}
\caption{\textit{Left-hand panels}: Inferred PM values of M31 background QSOs in right ascension (\(\mu_{\alpha}^{*}\); left) and declination (\(\mu_{\delta}\); right) as a function of \(G_{BP}-G_{RP}\) color.  
The upper sub-panels show the inferred PMs in each color bin for the full QSO sample (gray points), with the dashed gray line and shaded region indicating the overall zero-point offset and its \(1\sigma\) uncertainty from the MCMC analysis. The lower sub-panels show results for QSOs with different astrometric solutions: blue and orange points correspond to the inferred PMs in each color bin for 5p and 6p sources, respectively. Dashed lines and shaded areas denote the overall PMs and their  \(1\sigma\) uncertainties for the two subsamples.  
\textit{Right-hand panels}: Same as the \textit{left-hand panels}, but for M33 background QSOs. }
\label{fig:qso_zpt}
\end{figure*}

Previous studies have shown that zero-point offsets still exist in {\it Gaia} PM measurements (\citealt{Fabricius2021,Salomon2021,Rusterucci2024}). 
Building on the discussion above in Sect.~\ref{subsec:origin of discrepancy}, we used QSOs with both 5p and 6p astrometric solutions, whose PMs are expected to be zero due to their extragalactic nature, to determine the corresponding zero-point offsets in the directions of M31 and M33 for each solution type, and to assess whether systematic differences exist between the 5p and 6p solutions. The QSO sample was constructed by crossmatching sources from the {\it Gaia} DR3 \texttt{in\_qso\_candidates} table with the QSO catalog compiled by \citet{Liao2019}. This crossmatch serves to ensure the purity of the selected QSO sample. To further ensure the statistical reliability of the 6p solution analysis, we supplemented the QSO sample with additional {\it Gaia} QSO candidates flagged with \texttt{astrometric\_selection\_flag == 1} for the 6p solution, which have an estimated purity of 98\% \citep{Gaia2023}.

For M31, we selected background QSOs within an angular radius of \(10^\circ\) from the galaxy center. This selection was motivated by the \(5\)–\(10~{\rm \mu as\,yr^{-1}}\) zero-point variations reported over angular scales of \(4^\circ\)–\(20^\circ\) by \citet{Salomon2021}. Limiting the sample to within \(10^\circ\) reduces spatial systematics while preserving a sufficient number of QSOs for a robust zero-point estimate.
\begin{enumerate}
\item $\omega - 2\sigma_{\omega} < 0$;
\item $\mu_{\alpha}^{*} - \sigma_{\mu_{\alpha}^{*}} < 0 < \mu_{\alpha}^{*} + \sigma_{\mu_{\alpha}^{*}}$;
\item $\mu_{\delta} - \sigma_{\mu_{\delta}} < 0 < \mu_{\delta} + \sigma_{\mu_{\delta}}$;
\item RUWE $< 1.4$, \\
\texttt{ipd\_frac\_multi\_peak} $<$ 2,\\
\texttt{ipd\_gof\_harmonic\_amplitude} $<$ 0.1 ;
\item \texttt{astrometric\_matched\_observations} $\geq 10$.
\end{enumerate}

After applying these criteria, 1,670 QSOs with a 5p solution and 931 with a 6p solution remain.
Considering the spatial dependence of zero-point offsets noted earlier (e.g., \citealt{Salomon2021}; \citealt{Rusterucci2024}), typically on the order of \(5\)–\(10~{\rm \mu as\,yr^{-1}}\), these variations cannot explain the much larger PM discrepancy of about \(50~{\rm \mu as\,yr^{-1}}\) observed between the blue and red samples \citep{Rusterucci2024}. We therefore examined the color dependence of the zero-point offsets as a function of \(G_{BP} - G_{RP}\). To ensure statistical robustness, the full sample of 2,601 sources was divided into 13 color bins, each containing at least 200 sources and spanning a minimum color width of 0.04 mag.  
We inferred the zero-point offsets in each bin using the method described in Sect.~\ref{subsec:mcmc}, assuming the QSOs have no intrinsic motion, i.e., \(\boldsymbol{\mu}^{\rm model} = \boldsymbol{\mu}_{\text{sys}}\). Under this assumption, we applied Eq.~\ref{Eq:likelihood} to derive the posterior distributions in each color bin. The results, shown as gray symbols in the left-hand panels of Fig.~\ref{fig:qso_zpt}, indicate no significant color-dependent trend within the relatively narrow color range of the QSO sample, which spans from the red end of the blue supergiant sample to the blue end of the red supergiant sample.  
For the entire unbinned QSO sample, the overall zero-point offsets are $(\mu_{\alpha}^{*}, \mu_{\delta})_{\rm zpt}^{\rm M31} =
(-2.6 \pm 5.1,\ -2.7 \pm 4.1)~{\rm \mu as\,yr^{-1}}$, 
estimated using the same posterior inference method, and are largely consistent  
with those reported previously \citep{Salomon2021, Rusterucci2024}.

Then, using the same sample, we separately examined the potential color dependence of the zero-point offsets for the 5p and 6p QSO subsamples, and assessed whether any systematic differences exist between the two solution types. 
Following the same approach as above, the sources were binned by $G_{BP} - G_{RP}$ color, with each bin containing at least 150 sources for the 5p sample and 100 for the 6p sample, and a minimum color width of 0.04~mag.  
The zero-point offsets in each bin were inferred using the same method described previously and are shown as blue (5p) and orange (6p) points in the left-hand panels of Fig.~\ref{fig:qso_zpt}.  
No clear color-dependent trend is observed within either solution type; however, the 5p and 6p sources exhibit notably different zero-point offsets. The offsets for the 6p solution are substantially larger in magnitude and have larger uncertainties compared to those of the 5p solution.  
The overall zero-point offsets for each solution type were then derived using the posterior inference method, yielding 
$(\mu_{\alpha}^{*}, \mu_{\delta})_{\rm zpt,\ 5p}^{\rm M31} =
(-1.5 \pm 5.4,\ -2.1 \pm 4.3)\,{\rm \mu as\,yr^{-1}}$ and  
$(\mu_{\alpha}^{*}, \mu_{\delta})_{\rm zpt,\ 6p}^{\rm M31} =
(-11.5 \pm 16.8,\ -6.8 \pm 13.0)\,{\rm \mu as\,yr^{-1}}$.  
As was expected, the uncertainties in the 6p solution are approximately three to four times larger than those of the 5p solution.

For M33, we applied a similar procedure to determine the zero-point offsets, using approximately 7,000 5p and 3,800 6p background QSO candidates located within \(10^\circ\) from M33 center. Following the same cleaning process as for M31, 1,961 QSOs with a 5p solution and 840 with a 6p solution were finally selected. 
Similar to M31, we first examined whether the zero-point offsets for M33 are dependent on \(G_{BP} - G_{RP}\) color. Accordingly, the full sample of 2,801 sources was divided into 14 bins, each containing more than 200 sources ranging in magnitude by over 0.04 mag. The inferred PM values in each bin are shown as the gray symbols in the right-hand panels of Fig.~\ref{fig:qso_zpt}. As with M31, no significant color-dependent trend is observed across the QSO sample’s color range for M33. The overall zero-point offsets were derived as $(\mu_{\alpha}^{*}, \mu_{\delta})_{\rm zpt}^{\rm M33} =
(2.8 \pm 6.4,\ 2.0 \pm 5.3)\,{\rm \mu as\,yr^{-1}}$.
We then derived the zero-point offsets separately for the 5p and 6p subsamples, with the results again presented in the right-hand panels of Fig.~\ref{fig:qso_zpt}.
The pattern closely resembles that of M31: both the 5p and 6p sources exhibit nonzero, color-independent PM zero-points, but the offsets differ significantly, with the 6p solution generally showing larger offsets and greater uncertainties. For M33, the inferred overall zero-point offsets for the 5p and 6p solutions are: $(\mu_{\alpha}^{*}, \mu_\delta)_{\rm zpt,\ 5p}^{\rm M33} = (4.4 \pm 6.7,\ 2.6 \pm 5.5)\ \mu\mathrm{as}\ \mathrm{yr}^{-1}$ and $(\mu_{\alpha}^{*}, \mu_\delta)_{\rm zpt,\ 6p}^{\rm M33} = (-17.8 \pm 23.5,\ -7.9 \pm 20.2)\ \mu\mathrm{as}\ \mathrm{yr}^{-1}$, respectively. Again, the uncertainties in the 6p solution are approximately three to four times larger than those of the 5p solution.

Based on the above analysis, we find no significant color dependence in the zero-point offsets. However, a substantial systematic difference exists between the zero-points of the 5p and 6p solutions, which leads to the observed PM discrepancies between the blue and red samples, given their significantly different 5p and 6p fractions. Therefore, it is necessary to apply separate zero-point corrections for sources of each solution type. This correction will help further reduce the PM discrepancies observed between the blue and red samples in M31, as indicated below.

\subsection{Proper motion of M31} \label{subsec:M31PM}
\begin{figure*}[ht!]
\centering
\includegraphics[width=\textwidth]{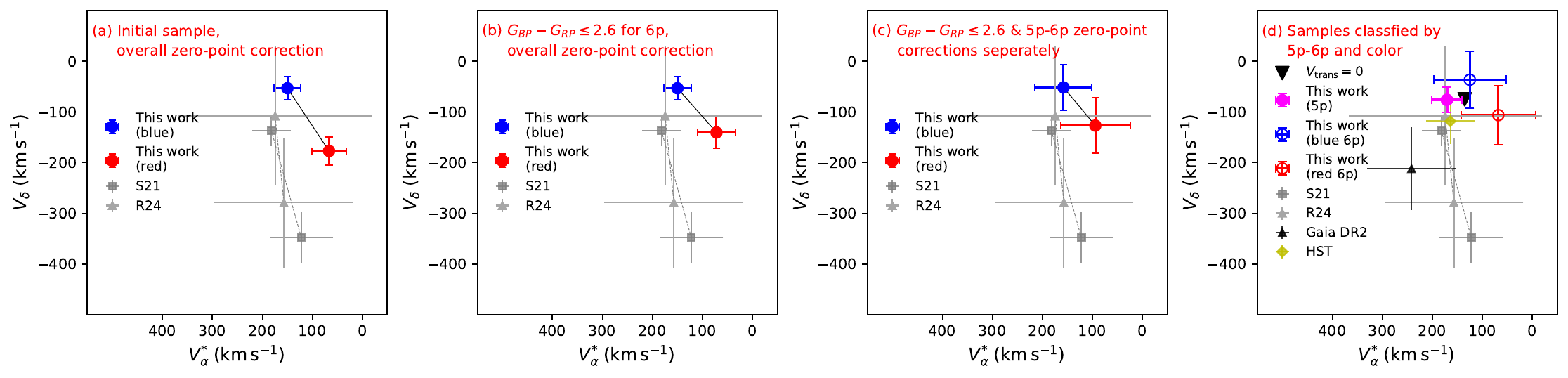}
\caption{Procedure for addressing the M31 PM discrepancy measured between the blue and red samples in the heliocentric frame assuming a distance of 784 kpc taken from \citet{Stanek1998}. Each panel is labeled in the topleft corner to indicate the corresponding sample. 
In Panels~(a)–(c), our PM results from the blue and red samples are connected by solid black lines. For comparison, we also show two other measurements based on {\it Gaia} (E)DR3 from \citet{Salomon2021} (gray squares) and \citet{Rusterucci2024} (gray triangles), connected by dashed gray lines.
In Panel~(d), we compare our final results with previous measurements. 
Here, the black inverted triangle indicates a purely radial orbit of M31 relative to the MW. 
The filled magenta and open blue and red points show our newly results from 5p, 6p blue, and 6p red samples ($G_{BP}-G_{RP} \leq 2.6$), respectively. In addition to {\it Gaia} (E)DR3, other measurements include results from HST observations \citep{Sohn2012, vdm2012a} (olive diamond) and {\it Gaia} DR2 \citep{vdm2019} (black triangle).}
    \label{fig:M31_result}
\end{figure*}

Based on the above discussion, we now present our approach to resolving the discrepancy in M31 PM measurements between the blue and red samples, which arises from two issues associated with the 6p solution.
We first revisited the PM results derived from the initial samples constructed in Sect.~\ref{subsec:M31RC}, as is shown in Fig.~\ref{fig:M31_result}. Following \citet{Salomon2021}, we applied the overall zero-point correction $(\mu_{\alpha}^{*}, \mu_{\delta})_{\rm zpt}^{\rm M31}$, as determined in Sect.~\ref{subsec:zero-point}, to both the red and blue samples. As is clearly shown in Panel~(a) of Fig.~\ref{fig:M31_result}, a significant discrepancy persists between the two samples, in agreement with previous results of \citet{Salomon2021} and \citet{Rusterucci2024}.
To address this issue, and guided by the tests presented in Sect.~\ref{subsec:origin of discrepancy}, we first removed extremely red stars with $G_{BP} - G_{RP} > 2.6$ from the red 6p sample, as their PM measurements are likely unreliable. 
After this cut, the updated PM measurements from both the blue and red samples, shown in Panel~(b) of Fig.~\ref{fig:M31_result}, are consistent within $2\sigma$ uncertainties, significantly reducing the discrepancy observed in the initial samples.
Building on this color cut, we further derived the M31 PM values by applying zero-point corrections separately to the 5p and 6p solutions, using background QSOs matched to each solution type, as described in Sect.~\ref{subsec:zero-point}. The resulting measurements are shown in Panel~(c) of Fig.~\ref{fig:M31_result}, where the PM values from the blue and red samples agree within $1\sigma$ uncertainties.
Following the two-step refinement, the tension in M31 PM measurements between the blue and red samples is effectively resolved.

Finally, we examined the results using both color and solution type: 5p (261 stars), blue 6p (551 stars), and red 6p stars with $G_{BP} - G_{RP} \leq 2.6$ (620 stars). The 5p sample is not further divided by color, as red 5p sources are too scarce for meaningful analysis.
After applying the zero-point corrections separately for the 5p and 6p solutions, we determined the corrected systemic PM and the corresponding transverse velocity in the heliocentric frame. These results are presented in the first three rows of Table~\ref{tab:result_M31} and shown in Panel~(d) of Fig.~\ref{fig:M31_result}.
Overall, the results from the three subsamples are consistent within $1\sigma$, further demonstrating the effectiveness of our procedure in resolving the PM discrepancy between the blue and red samples. Nevertheless, the larger uncertainties associated with the 6p results compared to the 5p measurement suggest that the 5p sample, despite having fewer sources, yields more accurate and robust result due to its significantly smaller uncertainties. Compared with previous measurements, as also shown in Panel~(d), our 5p result is in excellent agreement with that obtained from HST \citep{Sohn2012, vdm2012a}, but offers improved precision. 

\renewcommand{\arraystretch}{1.15} 
\begin{table*}[htbp]
\centering
\caption{Results for PM and the corresponding transverse velocity of the M31 COM derived from 5p solution sample, blue 6p solution sample and red 6p solution sample with $G_{BP}-G_{RP}\leq2.6$.\label{tab:result_M31}}
{\fontsize{10}{10}\selectfont
\begin{tabular}{ccccccc}
\hline
Sample (\#) & $\mu_{\alpha}^{*}$ $\rm(\mu as\,yr^{-1})$ &  $\mu_{\delta}$ $\rm(\mu as\,yr^{-1})$ & $\mu$ $\rm(\mu as\,yr^{-1})$ & $V_{\alpha*}$ $\rm(km\,s^{-1})$  & $V_{\delta}$ $\rm(km\,s^{-1})$ & $V_{\rm trans}$ $\rm(km\,s^{-1})$ \\
\hline
\multicolumn{7}{c}{Heliocentric frame} \\
5p (261) & $45.9 \pm 8.1$ & $-20.5 \pm 6.6$ & $50.8 \pm 7.8$ & $170.7 \pm 30.0$ & $-76.0 \pm 24.6$ & $188.7 \pm 28.9$ \\ 
6p Blue (551) & 33.6 $\pm$ 19.4 & $-$9.8 $\pm$ 15.1 & 39.2 $\pm$ 16.9 & 125.0 $\pm$ 72.0 & $-$36.5 $\pm$ 56.1 & 145.6 $\pm$ 63.0 \\ 
6p Red (620) & 18.3 $\pm$ 20.0 & $-$28.6 $\pm$ 15.6 & 39.3 $\pm$ 15.6 & 68.0 $\pm$ 74.3 & $-$106.3 $\pm$ 57.8 & 146.1 $\pm$ 57.9 \\ 
\hline
\multicolumn{7}{c}{Galactocentric rest frame} \\
5p (261)& 9.5 $\pm$ 8.1 & $-$0.3 $\pm$ 6.6 & 12.6 $\pm$ 6.4 & 35.4 $\pm$ 30.0 & $-$0.9 $\pm$ 24.6 & 46.7 $\pm$ 24.0\\ 
6p Blue (551) & $-$2.8 $\pm$ 19.4 & 10.4 $\pm$ 15.1 & 23.8 $\pm$ 12.4 & $-$10.3 $\pm$ 72.0 & 38.6 $\pm$ 56.1 & 88.5 $\pm$ 45.9 \\ 
6p Red (620) & $-$18.1 $\pm$ 20.0 & $-$8.4 $\pm$ 15.6 & 28.5 $\pm$ 14.7 & $-$67.3 $\pm$ 74.3 & $-$31.2 $\pm$ 57.8 & 105.9 $\pm$ 54.5 \\ 
\hline
\end{tabular}
\tablefoot{All results are corrected for zero-point offsets calculated from background QSOs, separately for the 5p and 6p solutions, with the uncertainties calculated as the root‐mean‐square of the formal MCMC uncertainty and the systematic uncertainty from the zero-point correction. The total PM $\mu$ is derived via 10,000 Monte Carlo sampling by propagating the uncertainties in $\mu_{\alpha}^{*}$ and $\mu_\delta$.}
}
\end{table*}

\begin{figure*}[htbp!]
\centering
\begin{minipage}{0.475\textwidth}
	\includegraphics[width=7.5cm,height = 7cm]{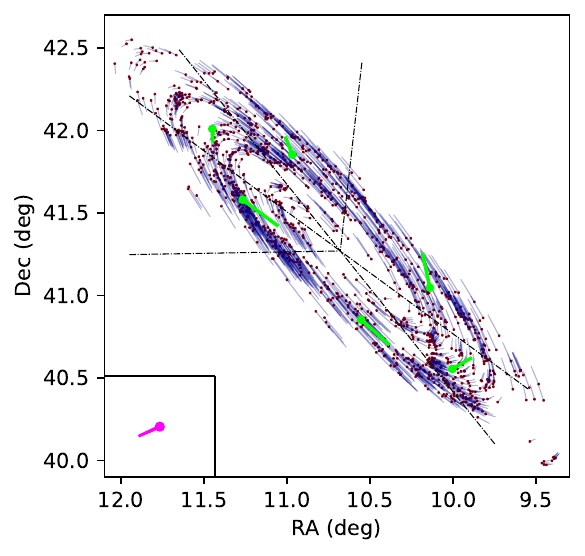}
	\end{minipage}
	\begin{minipage}{0.475\textwidth}
	\includegraphics[width=7.5cm,height = 7cm]{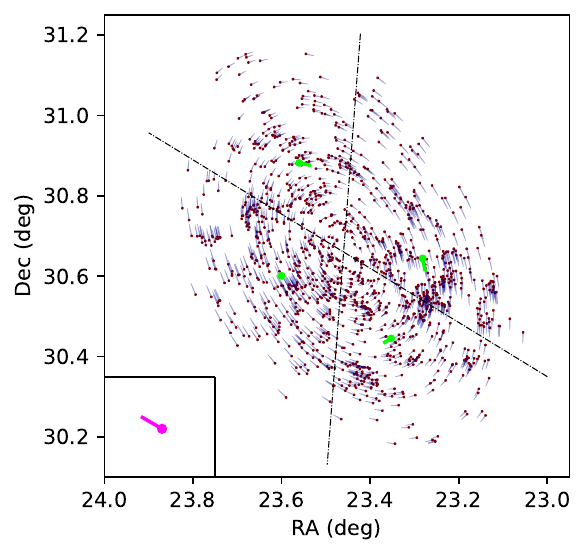}
	\end{minipage}
    \caption{PM kinematics of M31 (left) and M33 (right). All 5p and 6p sample stars with $G_{BP}-G_{RP} \leq 2.6$ are marked as red dots. The blue segments represent their predicted PMs from disk rotation, as determined in Sects.~\ref{subsec:M31RC} and~\ref{subsec:M33RC}, without considering the systemic PM component. The derived M31 and M33 COM PMs (after zero-point correction) using the 5p sample in the heliocentric frame are shown as the magenta segments at the bottomleft corner of each panel. Green segments illustrate the average observed PMs in six (four) position angle bins for M31 (M33), after zero-point correction and subtraction of the COM PM. Each bin contains more than 210 (300) sources for M31 (M33), with bin boundaries indicated by dashed black lines. In each case, the PM direction starts at the dot and points along the line segment.}
    \label{fig:M31_M33_segment}
\end{figure*}

After deriving the results, we provide a more intuitive illustration of the PM kinematic of M31 sample stars and its COM in the left panel of Fig.~\ref{fig:M31_M33_segment}. 
The full sample consists of 1,432 sources (5p stars and 6p stars with $G_{BP} - G_{RP} \leq 2.6$), for which the observed PMs were corrected for zero-point offsets separately according to their solution type.
The tracers were then divided into six bins in position angle, each containing at least 210 sources. The differences between the average observed PMs and the systemic PM derived from the 5p sample (first row in Table~\ref{tab:result_M31}, denoted by the magenta segment at the left panel of Fig.~\ref{fig:M31_M33_segment}) for these six bins exhibit a clear circular motion pattern across the disk, consistent with the predictions from the RC. This agreement further validates the robustness of our results.

To transform the derived PMs and transverse velocity from the heliocentric frame to the galactocentric rest frame, and to determine the relative motion between M31 and the MW, we must correct for the solar reflex motion. In this work, we used the solar motion of $(U_{\odot},V_{\odot},W_{\odot})\,=\,(7.01,\ 10.13,\ 4.95)\,\rm km\,s^{-1}$ \citep{Huang2015} and the circular speed at the solar position $V_{\rm c}(R_0)\,=$ 234.04 $\rm km\,s^{-1}$ \citep{Zhou2023}. This leads to a reflex motion of $(\mu_{\alpha}^{*},\mu_{\delta})_{\odot} = (36.4, -20.2)\,{\rm \mu as\,yr^{-1}}$ for M31 at 784~kpc, represented by the black inverted triangle in Panel (d) of Fig.~\ref{fig:M31_result}. If the derived PMs in the heliocentric frame match these values, it indicates zero transverse and a purely radial motion between M31 and the MW. After correcting for this solar reflex motion, we derived the PMs and corresponding transverse velocities in the galactocentric rest frame, which are provided in the last three rows of Table~\ref{tab:result_M31}. The result from the 5p sample corresponds to a transverse velocity of $V_{\rm trans} = 46.7 \pm 24.0 \, \rm km\,s^{-1}$, in agreement with recent cosmological simulations \citep{Sawala2023}. This supports the scenario of a nearly radial orbit for the relative motion between M31 and the MW, suggesting that the LG is gravitationally bound.

By combining the PM measurement from the 5p sample with the systemic LOS velocity of M31 derived in Sect.~\ref{subsec:M31RC}, both corrected for solar reflex motions, we derived the galactocentric 3D velocity of the M31 COM, $\vec{v}_{\text{DR3}}^{\rm M31} = (V_{X}, V_{Y}, V_{Z}) = (24.1 \pm 25.9,\ -107.9 \pm 17.4,\ 38.8 \pm 22.8)\,\rm km\,s^{-1}$.
The uncertainties were propagated from the individual components using a Monte Carlo method.

Notably, the CMD distributions of our blue and red samples differ slightly from those used by \citet{Salomon2021}. To assess the robustness of our results to sample selection, we reanalyzed an alternative CMD-selected sample following \citet{Salomon2021}, applying the same quality cuts described in Sects.~\ref{subsec:cleaning} and~\ref{subsec:M31RC}, together with our correction procedure. The PM results at each step show excellent agreement with those from our supergiant samples, confirming the robustness of our measurements and the reliability of our methodology in addressing the discrepancy between blue and red samples.

\subsection{Proper motion of M33} \label{subsec:M33PM}

\renewcommand{\arraystretch}{1.15} 
\begin{table*}[htbp!]
\footnotesize
\centering
\caption{Results for PM and the corresponding transverse velocity of the M33 COM derived from 5p solution sample, blue 6p solution sample and red 6p solution sample with $G_{BP}-G_{RP}\leq2.6$. \label{tab:result_M33}}
{\fontsize{10}{10}\selectfont
\begin{tabular}{ccccccc}
\hline
Sample (\#) & $\mu_{\alpha}^{*}$ $\rm(\mu as\,yr^{-1})$ &  $\mu_{\delta}$ $\rm(\mu as\,yr^{-1})$ & $\mu$ $\rm(\mu as\,yr^{-1})$ & $V_{\alpha*}$ $\rm(km\,s^{-1})$  & $V_{\delta}$ $\rm(km\,s^{-1})$ & $V_{\rm trans}$ $\rm(km\,s^{-1})$ \\
\hline
\multicolumn{7}{c}{Heliocentric frame} \\
5p (334) & 45.3 $\pm$ 9.7 & 26.3 $\pm$ 7.3 & 52.9 $\pm$ 9.1 & 180.2 $\pm$ 38.8 & 104.6 $\pm$ 29.0 & 210.8 $\pm$ 36.3 \\ 
6p Blue (502) & 49.9 $\pm$ 26.9 & 4.5 $\pm$ 22.1 & 55.9 $\pm$ 24.1 & 198.7 $\pm$ 107.2 & 18.1 $\pm$ 87.9 & 222.7 $\pm$ 95.9 \\ 
6p Red (570) & 25.1 $\pm$ 26.9 & 26.0 $\pm$ 22.0 & 45.7 $\pm$ 20.8 & 100.0 $\pm$ 107.1 & 103.5 $\pm$ 87.5 & 181.8 $\pm$ 82.8 \\ 
\hline
\multicolumn{7}{c}{Galactocentric rest frame} \\
5p (334)& 7.1 $\pm$ 9.7 & 57.7 $\pm$ 7.3 & 58.9 $\pm$ 7.3 & 28.1 $\pm$ 38.8 & 229.6 $\pm$ 29.0 & 234.5 $\pm$ 29.2\\ 
6p Blue (502) & 11.7 $\pm$ 26.9 & 35.9 $\pm$ 22.1 & 47.2 $\pm$ 20.3 & 46.6 $\pm$ 107.2 & 143.1 $\pm$ 87.9 & 187.8 $\pm$ 80.9 \\ 
6p Red (570) & $-$13.1 $\pm$ 26.9 & 57.4 $\pm$ 22.0 & 65.0 $\pm$ 21.4 & $-$52.1 $\pm$ 107.1 & 228.5 $\pm$ 87.5 & 258.8 $\pm$ 85.3 \\ 
\hline
\end{tabular}
\tablefoot{All results are corrected for zero-point offsets calculated from background QSOs, separately for the 5p and 6p solutions, with the uncertainties calculated as the root‐mean‐square of the formal MCMC uncertainty and the systematic uncertainty from the zero-point correction. The total PM $\mu$ is derived via 10,000 Monte Carlo sampling by propagating the uncertainties in $\mu_{\alpha}^{*}$ and $\mu_\delta$.}}
\end{table*}

\begin{figure*}[htbp!]
\centering
\includegraphics[width=\textwidth]{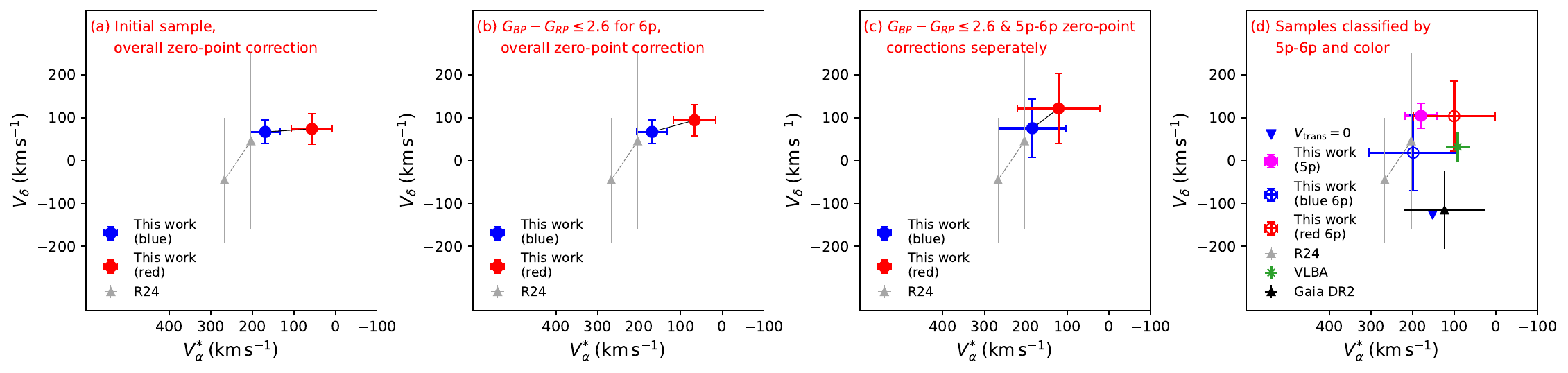}
\caption{Procedure for determining M33's systemic PM in the heliocentric frame, assuming a distance of 840 kpc \citep{Galleti2004distance,Breuval2023}. Each panel is labeled in the topleft corner to indicate the corresponding sample. The results at each step are indicated using the same symbols as in Fig.~\ref{fig:M31_result}. For comparison, results from \citet{Rusterucci2024} based on {\it Gaia} DR3 are shown as gray triangles connected by dashed gray lines. In Panel~(d), we compare our final results with previous measurements. The blue inverted triangle indicates a purely radial orbit of M33 relative to the MW. The filled magenta and open blue and red points show our newly results from 5p, 6p blue, and 6p red samples ($G_{BP}-G_{RP} \leq 2.6$), respectively. The green cross represents the measurement by VLBA \citep{Brunthaler2005} and the black triangle shows the result from {\it Gaia} DR2 \citep{vdm2019}.}
\label{fig:M33_result}
\end{figure*}

Following a similar procedure as for M31, we next derived the systemic PM of M33 in the heliocentric frame.  
We began by remeasuring the PM using the initial sample constructed in Sect.~\ref{subsec:M33RC}, applying the overall zero-point correction $(\mu_{\alpha}^{*},\mu_{\delta})_{\rm zpt}^{\rm M33}$ derived in Sect.~\ref{subsec:zero-point}. The results are shown in Panel (a) of Fig.~\ref{fig:M33_result}.
The discrepancy between the blue and red samples is notably smaller than that observed in M31, particularly in the declination component, consistent with the findings of \citet{Rusterucci2024}. This is primarily due to the limited number of extremely red supergiants with $G_{BP}-G_{RP} > 2.6$ in M33; that is, the PM measurements in M33 are less affected by the first issue of the 6p solution found in Sect.~\ref{subsec:origin of discrepancy}.
For the right ascension component, the observed difference between the blue and red samples mainly arises from the systematic offsets between the 5p and 6p astrometric solutions, further amplified by the differing proportions of 5p and 6p stars in the two samples.  
Similar to M31, we first excluded these extremely red 6p stars, and the corresponding results are shown in Panel~(b) of Fig.~\ref{fig:M33_result}.
On top of this color cut, we further derived the systemic PM of M33 by applying separate zero-point corrections to the 5p and 6p stars, using background QSOs corresponding to each solution type, as described in Sect.~\ref{subsec:zero-point}. These results are shown in Panel (c) of Fig.~\ref{fig:M33_result}. Clearly, this step effectively mitigates the discrepancy between the blue and red samples in the right ascension component observed in Panel (a).
Finally, we determined the M33 COM PM using three subsamples: 5p (334 stars), blue 6p (502 stars), and red 6p stars with $G_{BP}-G_{RP} \leq 2.6$ (570 stars). After applying the respective zero-point corrections for the 5p and 6p solutions, the final results are listed in the first three rows of Table~\ref{tab:result_M33} and illustrated in Panel (d) of Fig.~\ref{fig:M33_result}. Overall, the three measurements are consistent within $1\sigma$, with the 5p result being the most robust and accurate due to its smaller uncertainties. Compared to previous studies, our 5p measurement agrees with the VLBA result within $1.5\sigma$, with comparable uncertainties. 

The right panel of Fig.~\ref{fig:M31_M33_segment} displays a more intuitive PM kinematics of M33 sample stars and its COM. After applying zero-point corrections separately for 5p and 6p solutions, all 1,406 sources (5p stars and 6p stars with $G_{BP}-G_{RP} \leq 2.6$) were grouped into four position angle bins of over 300 sources each. The differences between the average observed PM values and the systemic PM derived from the 5p sample (first row in Table~\ref{tab:result_M33}, denoted by the magenta segment at the right panel of Fig.~\ref{fig:M31_M33_segment}) in these bins, exhibit a clear circular motion pattern. This aligns well with the predictions from the RC, thereby further confirming the reliability of our measurement.

To transform the derived PMs and transverse velocity from the heliocentric frame to the galactocentric rest frame, we then corrected for the solar reflex motion. Using the same solar motion parameters adopted for M31, we applied a reflex motion of $(\mu_{\alpha}^{*},\mu_{\delta})_{\odot}$ = (38.2, $-$31.4) ${\rm \mu as\,yr^{-1}}$ at the position of the M33 COM by adopting its distance of 840 kpc. After correcting for this solar reflx motion, the results are given in the last three rows of Table~\ref{tab:result_M33}. Combined with the M33 systemic LOS velocity derived in Sect.~\ref{subsec:M33RC}, we obtained the galactocentric 3D velocity $\vec{v}_{\text{DR3}}^{\rm M33}\,=\,(-50.3\pm30.6,\ 76.3\pm27.6,\ 215.4\pm24.5)\,\rm km\,s^{-1}$, with uncertainties estimated following the same procedure as in Sect.~\ref{subsec:M31PM}.

Based on the galactocentric 3D velocities of the M31 and M33 COM derived in this work, the relative motion between M33 and M31 can be described by a radial component of $V_{\text{rad, DR3}}\,=\,-169.7\,\rm km\,s^{-1}$ and a transverse component of $V_{\text{trans, DR3}}\,=\,204.6\,\rm km\,s^{-1}$. For M33, if we combine the result in this work (5p sample) with that from VLBA \citep{Brunthaler2005}, a weighted average PM result in the heliocentric frame can be derived as $(\mu_{\alpha}^{*},\mu_{\delta})_{\text{DR3+VLBA}}^{\rm M33}\,=\,(30.6\pm5.7,\ 19.0\pm5.7)\,\rm \mu as\,yr^{-1}$. Correcting for the solar reflex motion, this corresponds to a galactocentric 3D velocity $\vec{v}_{\text{DR3+VLBA}}^{\rm M33}\,=\,(1.5\pm18.5,\ 97.4\pm17.9,\ 181.8\pm19.6)\,\rm km\,s^{-1}$. The relative motion between M33 and M31 then comprises a radial component of $V_{\text{rad, DR3+VLBA}}\,=\,-185.0\,\rm km\,s^{-1}$ and a transverse component of $V_{\text{trans, DR3+VLBA}}\,=\,170.0\,\rm km\,s^{-1}$.

\begin{figure*}[ht!]
\centering
    \begin{minipage}{0.49\textwidth}
    \centering
    \includegraphics[width=\textwidth]{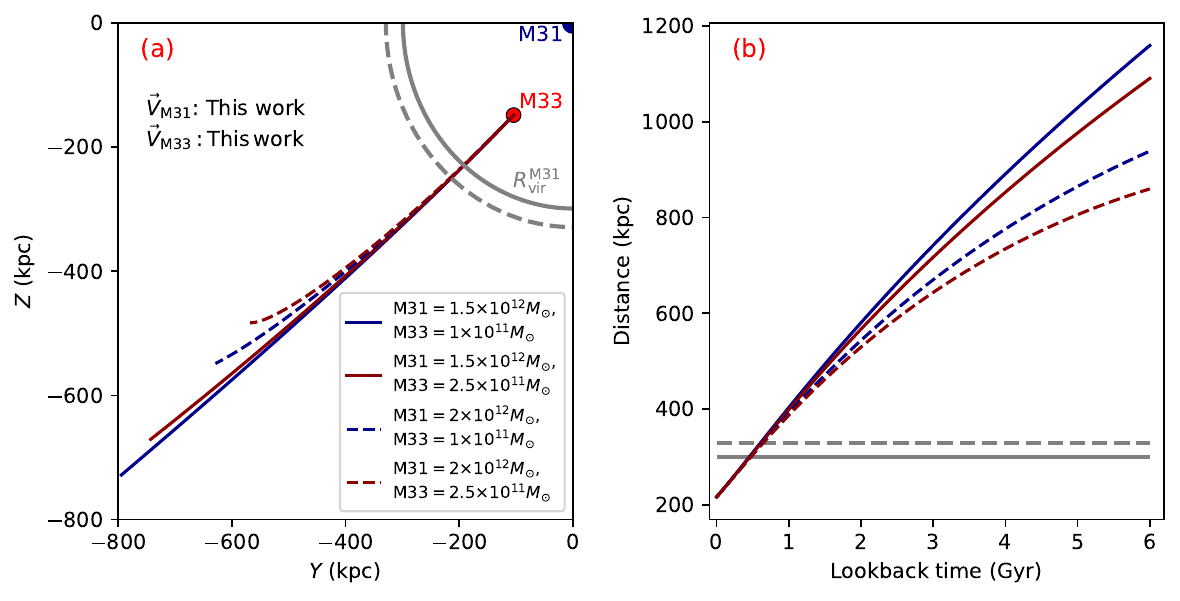}
    \end{minipage}
    \hspace{0.01\textwidth}
    \begin{minipage}{0.49\textwidth}
    \centering
    \includegraphics[width=\textwidth]{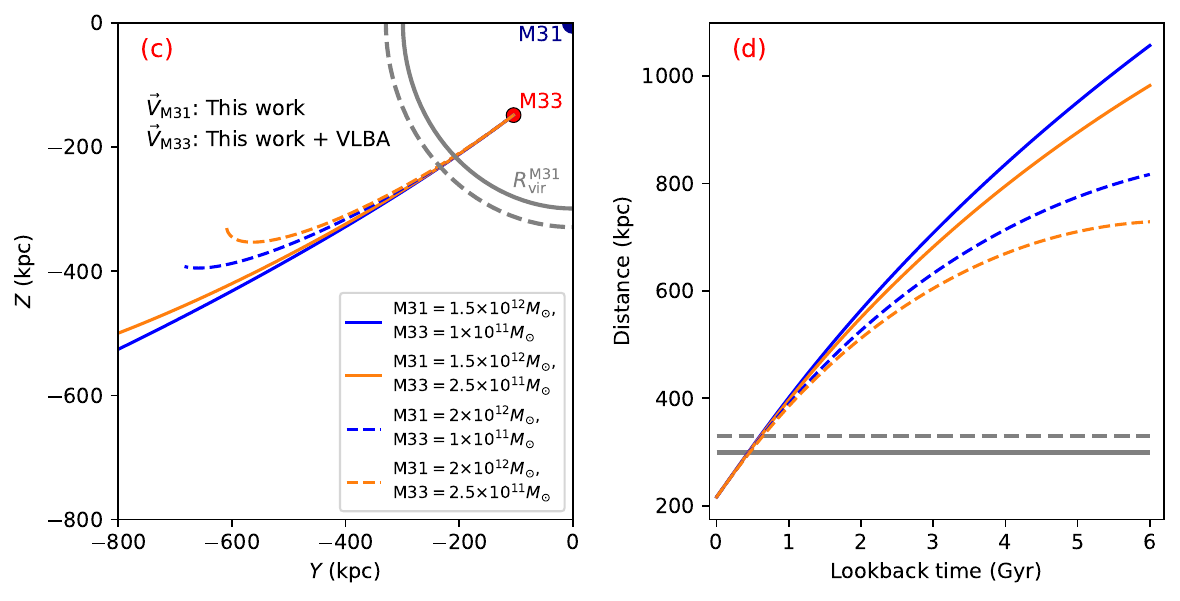}
    \end{minipage}
\caption{Numerical orbital integration backward for M33 in the galactocentric frame centered on M31. The 3D galactocentric velocity of M31 is taken from this work. Orbits are integrated backward for 6 Gyrs from the present-day position and velocity. Panels (a) and (b) show the orbit of M33 using its 3D velocity measured solely in this work, while panels (c) and (d) adopt the velocity from this work combined with the VLBA measurement. Similar to \citet{Patel2017} and \citet{vdm2019}, we consider four mass combinations for M31 and M33, with two corresponding virial radii of M31: $R_{\rm vir} = 299$ kpc (solid gray lines) for an M31 mass of $1.5 \times 10^{12} M_{\odot}$, and $R_{\rm vir} = 329$ kpc (dashed gray lines) for an M31 mass of $2 \times 10^{12} M_{\odot}$. Panels (a) and (c) display the orbit of M33 projected onto the $Y$-$Z$ plane, while panels (b) and (d) show the evolution of M33's distance from M31 as a function of lookback time.}
\label{fig:m33_orbit}
\end{figure*}

To investigate whether a first infall scenario is supported for M33 given the newly derived 3D velocities of M31 and M33, we followed a similar procedure to that of \citet{Patel2017} and \citet{vdm2019}, numerically integrating M33's orbit backward over 6 Gyr while neglecting the influence of the MW. The adopted potential models and parameters were identical to those used in \citet{Patel2017} and \citet{vdm2019}. The distances to M31 and M33 were set to 784 and 840 kpc, respectively, and their velocities were updated using the values derived in this work. For M33, we considered two cases: using only our {\it Gaia} DR3-based result ($\vec{v}_{\text{DR3}}^{\text{M33}}$, shown in the left-hand panels of Fig.~\ref{fig:m33_orbit}) and combining it with the VLBA result ($\vec{v}_{\text{DR3+VLBA}}^{\text{M33}}$, shown in the right-hand panels of Fig.~\ref{fig:m33_orbit}). In both cases, regardless of the assumed mass of M31 and M33, the results consistently support a first infall scenario; namely, that M33 is currently falling into M31's virial radius for the first time.

\section{Summary}
\label{section:summary_discussion}
In this work, we revisited the systemic PMs of M31 and M33 using massive supergiant stars as tracers. These tracers have been selected from color-color or color--magnitude diagrams, excluding foreground contaminants based on kinematic and astrometric properties, making them purer than those selected solely from CMD of {\it Gaia} broadband photometry. To further ensure sample purity, we applied several criteria to retain reliable disk objects.
To assess the contribution of the supergiants' disk rotation to the observed PMs, we established RC models for the disk of M31 and M33 using sources with LOS velocity measurements from the reliable disk objects selected earlier. 

Using the astrometric measurements from {\it Gaia} DR3, we derived the systemic PMs for M31 and M33. For M31, the results from initial samples show significant discrepancy between blue and red samples, in line with previous studies. To address this, we identified its origin as arising from systematic astrometric offsets between sources with a 6p astrometric solution and those with a 5p one. The discrepancy was further amplified by the differing fractions of 5p and 6p sources in the blue and red samples. The red sample is dominated by 6p sources while the blue sample includes a substantial fraction of 5p sources. Specifically, we identified two main issues associated with 6p sources:
1) unreliable astrometric measurements for extremely red stars, specifically those with $G_{BP}-G_{RP} > 2.6$, which exhibit significant deviations from the bulk of the 6p population. Combined with the findings of \citet{Fabricius2021}, we conclude that the astrometry of these red 6p sources is likely unreliable;
2) a color-independent systematic offset between 5p and 6p solutions, which is caused by the different zero-point offsets between these two solution types.
To correct for these effects, we applied the following steps: 1) excluding 6p sources with extreme red colors by restricting the sample to those with $G_{BP} - G_{RP} \leq 2.6$; 2) correcting the PM zero-point offsets separately for 5p and 6p sources using background QSOs, thereby accounting for systematic differences between the two solution types.
After applying these corrections, the PM measurements derived from M31 blue and red samples become consistent within 1$\sigma$, and both indicate a radially dominated orbit of M31 relative to the MW. This resolves the previously found discrepancy between PM measurements using blue and red samples and provides robust evidence of a radially dominated M31–MW relative motion.

Using 5p sample stars, which provide the most accurate and robust result, we derived a galactocentric transverse velocity of $V_{\rm trans} = 46.7 \pm 24.0\,{\rm km\,s^{-1}}$ for M31. This is in excellent agreement with previous HST-based measurement, but with improved precision.
For M33, our measurement is consistent with previous VLBA measurements within 1.5$\sigma$. Combining both M31 and M33 measurements, M33's first infall scenario is supported after numerical integration, regardless of the mass model adopted.

Overall, we present the most robust {\it Gaia}-based PM measurements for M31 and M33 to date, conclusively demonstrating that the relative motion between M31 and the MW is radially dominated. 
Our work also highlights the systematic offsets between {\it Gaia} 5p and 6p astrometric solutions, as well as the biases observed in extreme red 6p sources. Given that approximately twothirds of the sources in {\it Gaia} DR3 lack 5p solution, especially significant for faint sources (the fraction of 5p sources decreases from 88\% at $G < 18$ to only 1.4\% at $20 < G < 21$; \citealt{Lindegren2021ast}), we recommend additional caution when using the 6p solution in cases requiring high-precision astrometry. This work also provides valuable insights that may help mitigate these issues in the upcoming {\it Gaia} DR4.

\begin{acknowledgements}
We are grateful to Prof. Haibo Yuan for his invaluable comments and insightful discussions.
This work acknowledges the supports from National Key R\&D Programme of China (Grant No. 2024YFA1611903) and the National Science Foundation of China (NSFC Grant No. 12422303, 12090040 and 12090044).\\
This work has made use of data from the European Space Agency (ESA) mission {\it Gaia} (\url{https://www.cosmos.esa.int/gaia}), processed by the {\it Gaia} Data Processing and Analysis Consortium (DPAC, \url{https:// www.cosmos.esa.int/web/gaia/dpac/ consortium}). Funding for the DPAC has been provided by national institutions, in particular the institutions participating in the {\it Gaia} Multilateral Agreement. \\

This work has made use of data products from the Guo Shou Jing Telescope (the Large Sky Area Multi-Object Fibre Spectroscopic Telescope, LAMOST). LAMOST is a National Major Scientific Project built by the Chinese Academy of Sciences. Funding for the project has been provided by the National Development and Reform Commission. LAMOST is operated and managed by the National Astronomical Observatories, Chinese Academy of Sciences.
\end{acknowledgements}

\bibliographystyle{aa}
\bibliography{reference}

\begin{thebibliography}{73}
\expandafter\ifx\csname natexlab\endcsname\relax\def\natexlab#1{#1}\fi

\bibitem[{{Azimlu} {et~al.}(2011){Azimlu}, {Marciniak}, \& {Barmby}}]{Azimlu2011}
{Azimlu}, M., {Marciniak}, R., \& {Barmby}, P. 2011, \aj, 142, 139

\bibitem[{{Bekki}(2008)}]{Bekki2008}
{Bekki}, K. 2008, \mnras, 390, L24

\bibitem[{{Benisty} {et~al.}(2022){Benisty}, {Vasiliev}, {Evans}, {Davis}, {Hartl}, \& {Strigari}}]{Benisty2022}
{Benisty}, D., {Vasiliev}, E., {Evans}, N.~W., {et~al.} 2022, \apjl, 928, L5

\bibitem[{{Breuval} {et~al.}(2023){Breuval}, {Riess}, {Macri}, {Li}, {Yuan}, {Casertano}, {Konchady}, {Trahin}, {Durbin}, \& {Williams}}]{Breuval2023}
{Breuval}, L., {Riess}, A.~G., {Macri}, L.~M., {et~al.} 2023, \apj, 951, 118

\bibitem[{{Brunthaler} {et~al.}(2005){Brunthaler}, {Reid}, {Falcke}, {Greenhill}, \& {Henkel}}]{Brunthaler2005}
{Brunthaler}, A., {Reid}, M.~J., {Falcke}, H., {Greenhill}, L.~J., \& {Henkel}, C. 2005, Science, 307, 1440

\bibitem[{{Bullock} \& {Boylan-Kolchin}(2017)}]{Bullock2017}
{Bullock}, J.~S. \& {Boylan-Kolchin}, M. 2017, \araa, 55, 343

\bibitem[{{Carlesi} {et~al.}(2022){Carlesi}, {Hoffman}, \& {Libeskind}}]{Carlesi2022}
{Carlesi}, E., {Hoffman}, Y., \& {Libeskind}, N.~I. 2022, \mnras, 513, 2385

\bibitem[{{Chemin} {et~al.}(2009){Chemin}, {Carignan}, \& {Foster}}]{Chemin2009}
{Chemin}, L., {Carignan}, C., \& {Foster}, T. 2009, \apj, 705, 1395

\bibitem[{{Ciardullo} {et~al.}(2004){Ciardullo}, {Durrell}, {Laychak}, {Herrmann}, {Moody}, {Jacoby}, \& {Feldmeier}}]{Ciardullo2004}
{Ciardullo}, R., {Durrell}, P.~R., {Laychak}, M.~B., {et~al.} 2004, \apj, 614, 167

\bibitem[{{Corbelli} {et~al.}(2010){Corbelli}, {Lorenzoni}, {Walterbos}, {Braun}, \& {Thilker}}]{Corbelli2010}
{Corbelli}, E., {Lorenzoni}, S., {Walterbos}, R., {Braun}, R., \& {Thilker}, D. 2010, \aap, 511, A89

\bibitem[{{Corbelli} {et~al.}(2014){Corbelli}, {Thilker}, {Zibetti}, {Giovanardi}, \& {Salucci}}]{Corbelli2014}
{Corbelli}, E., {Thilker}, D., {Zibetti}, S., {Giovanardi}, C., \& {Salucci}, P. 2014, \aap, 572, A23

\bibitem[{{Cordiner} {et~al.}(2011){Cordiner}, {Cox}, {Evans}, {Trundle}, {Smith}, {Sarre}, \& {Gordon}}]{Cordiner2011ApJ}
{Cordiner}, M.~A., {Cox}, N. L.~J., {Evans}, C.~J., {et~al.} 2011, \apj, 726, 39

\bibitem[{{Cui} {et~al.}(2012){Cui}, {Zhao}, {Chu}, {Li}, {Li}, {Zhang}, {Su}, {Yao}, {Wang}, {Xing}, {Li}, {Zhu}, {Wang}, {Gu}, {Luo}, {Xu}, {Zhang}, {Liu}, {Zhang}, {Yang}, {Cao}, {Chen}, {Chen}, {Chen}, {Chen}, {Chu}, {Feng}, {Gong}, {Hou}, {Hu}, {Hu}, {Hu}, {Jia}, {Jiang}, {Jiang}, {Jiang}, {Jin}, {Li}, {Li}, {Li}, {Liu}, {Liu}, {Lu}, {Mao}, {Men}, {Qi}, {Qi}, {Shi}, {Tang}, {Tao}, {Wang}, {Wang}, {Wang}, {Wang}, {Wang}, {Wang}, {Wang}, {Wang}, {Wang}, {Wang}, {Wang}, {Wang}, {Xu}, {Xu}, {Yang}, {Yu}, {Yuan}, {Yuan}, {Zhai}, {Zhang}, {Zhang}, {Zhang}, {Zhao}, {Zhou}, {Zhou}, {Zhu}, \& {Zou}}]{Cui2012}
{Cui}, X.-Q., {Zhao}, Y.-H., {Chu}, Y.-Q., {et~al.} 2012, Research in Astronomy and Astrophysics, 12, 1197

\bibitem[{{de Vaucouleurs} {et~al.}(1991){de Vaucouleurs}, {de Vaucouleurs}, {Corwin}, {Buta}, {Paturel}, \& {Fouque}}]{deVaucouleurs1991}
{de Vaucouleurs}, G., {de Vaucouleurs}, A., {Corwin}, Jr., H.~G., {et~al.} 1991, {Third Reference Catalogue of Bright Galaxies}

\bibitem[{{Drout} {et~al.}(2012){Drout}, {Massey}, \& {Meynet}}]{Drout2012}
{Drout}, M.~R., {Massey}, P., \& {Meynet}, G. 2012, \apj, 750, 97

\bibitem[{{Drout} {et~al.}(2009){Drout}, {Massey}, {Meynet}, {Tokarz}, \& {Caldwell}}]{Drout2009}
{Drout}, M.~R., {Massey}, P., {Meynet}, G., {Tokarz}, S., \& {Caldwell}, N. 2009, \apj, 703, 441

\bibitem[{{Ekstr{\"o}m} {et~al.}(2012){Ekstr{\"o}m}, {Georgy}, {Eggenberger}, {Meynet}, {Mowlavi}, {Wyttenbach}, {Granada}, {Decressin}, {Hirschi}, {Frischknecht}, {Charbonnel}, \& {Maeder}}]{Ekstrom2012}
{Ekstr{\"o}m}, S., {Georgy}, C., {Eggenberger}, P., {et~al.} 2012, \aap, 537, A146

\bibitem[{{Evans} {et~al.}(2010){Evans}, {Primini}, {Glotfelty}, {Anderson}, {Bonaventura}, {Chen}, {Davis}, {Doe}, {Evans}, {Fabbiano}, {Galle}, {Gibbs}, {Grier}, {Hain}, {Hall}, {Harbo}, {He}, {Houck}, {Karovska}, {Kashyap}, {Lauer}, {McCollough}, {McDowell}, {Miller}, {Mitschang}, {Morgan}, {Mossman}, {Nichols}, {Nowak}, {Plummer}, {Refsdal}, {Rots}, {Siemiginowska}, {Sundheim}, {Tibbetts}, {Van Stone}, {Winkelman}, \& {Zografou}}]{Evans2010}
{Evans}, I.~N., {Primini}, F.~A., {Glotfelty}, K.~J., {et~al.} 2010, \apjs, 189, 37

\bibitem[{{Fabricius} {et~al.}(2021){Fabricius}, {Luri}, {Arenou}, {Babusiaux}, {Helmi}, {Muraveva}, {Reyl{\'e}}, {Spoto}, {Vallenari}, {Antoja}, {Balbinot}, {Barache}, {Bauchet}, {Bragaglia}, {Busonero}, {Cantat-Gaudin}, {Carrasco}, {Diakit{\'e}}, {Fabrizio}, {Figueras}, {Garcia-Gutierrez}, {Garofalo}, {Jordi}, {Kervella}, {Khanna}, {Leclerc}, {Licata}, {Lambert}, {Marrese}, {Masip}, {Ramos}, {Robichon}, {Robin}, {Romero-G{\'o}mez}, {Rubele}, \& {Weiler}}]{Fabricius2021}
{Fabricius}, C., {Luri}, X., {Arenou}, F., {et~al.} 2021, \aap, 649, A5

\bibitem[{{Gaia Collaboration} {et~al.}(2023){Gaia Collaboration}, {Bailer-Jones}, {Teyssier}, {Delchambre}, {Ducourant}, {Garabato}, {Hatzidimitriou}, {Klioner}, {Rimoldini}, {Bellas-Velidis}, {Carballo}, {Carnerero}, {Diener}, {Fouesneau}, {Galluccio}, {Gavras}, {Krone-Martins}, {Raiteri}, {Teixeira}, {Brown}, {Vallenari}, {Prusti}, {de Bruijne}, {Arenou}, {Babusiaux}, {Biermann}, {Creevey}, {Evans}, {Eyer}, {Guerra}, {Hutton}, {Jordi}, {Lammers}, {Lindegren}, {Luri}, {Mignard}, {Panem}, {Pourbaix}, {Randich}, {Sartoretti}, {Soubiran}, {Tanga}, {Walton}, {Bastian}, {Drimmel}, {Jansen}, {Katz}, {Lattanzi}, {van Leeuwen}, {Bakker}, {Cacciari}, {Casta{\~n}eda}, {De Angeli}, {Fabricius}, {Fr{\'e}mat}, {Guerrier}, {Heiter}, {Masana}, {Messineo}, {Mowlavi}, {Nicolas}, {Nienartowicz}, {Pailler}, {Panuzzo}, {Riclet}, {Roux}, {Seabroke}, {Sordo}, {Th{\'e}venin}, {Gracia-Abril}, {Portell}, {Altmann}, {Andrae}, {Audard}, {Benson}, {Berthier}, {Blomme}, {Burgess}, {Busonero}, {Busso}, {C{\'a}novas}, {Carry}, {Cellino},
  {Cheek}, {Clementini}, {Damerdji}, {Davidson}, {de Teodoro}, {Nu{\~n}ez Campos}, {Dell'Oro}, {Esquej}, {Fern{\'a}ndez-Hern{\'a}ndez}, {Fraile}, {Garc{\'\i}a-Lario}, {Gosset}, {Haigron}, {Halbwachs}, {Hambly}, {Harrison}, {Hern{\'a}ndez}, {Hestroffer}, {Hodgkin}, {Holl}, {Jan{\ss}en}, {Jevardat de Fombelle}, {Jordan}, {Lanzafame}, {L{\"o}ffler}, {Marchal}, {Marrese}, {Moitinho}, {Muinonen}, {Osborne}, {Pancino}, {Pauwels}, {Recio-Blanco}, {Reyl{\'e}}, {Riello}, {Roegiers}, {Rybizki}, {Sarro}, {Siopis}, {Smith}, {Sozzetti}, {Utrilla}, {van Leeuwen}, {Abbas}, {{\'A}brah{\'a}m}, {Abreu Aramburu}, {Aerts}, {Aguado}, {Ajaj}, {Aldea-Montero}, {Altavilla}, {{\'A}lvarez}, {Alves}, {Anderson}, {Anglada Varela}, {Antoja}, {Baines}, {Baker}, {Balaguer-N{\'u}{\~n}ez}, {Balbinot}, {Balog}, {Barache}, {Barbato}, {Barros}, {Barstow}, {Bartolom{\'e}}, {Bassilana}, {Bauchet}, {Becciani}, {Bellazzini}, {Berihuete}, {Bernet}, {Bertone}, {Bianchi}, {Binnenfeld}, {Blanco-Cuaresma}, {Boch}, {Bombrun}, {Bossini}, {Bouquillon},
  {Bragaglia}, {Bramante}, {Breedt}, {Bressan}, {Brouillet}, {Brugaletta}, {Bucciarelli}, {Burlacu}, {Butkevich}, {Buzzi}, {Caffau}, {Cancelliere}, {Cantat-Gaudin}, {Carlucci}, {Carrasco}, {Casamiquela}, {Castellani}, {Castro-Ginard}, {Chaoul}, {Charlot}, {Chemin}, {Chiaramida}, {Chiavassa}, {Chornay}, {Comoretto}, {Contursi}, {Cooper}, {Cornez}, {Cowell}, {Crifo}, {Cropper}, {Crosta}, {Crowley}, {Dafonte}, {Dapergolas}, {David}, \& {de Laverny}}]{Gaia2023}
{Gaia Collaboration}, {Bailer-Jones}, C.~A.~L., {Teyssier}, D., {et~al.} 2023, \aap, 674, A41

\bibitem[{{Gaia Collaboration} {et~al.}(2018){Gaia Collaboration}, {Brown}, {Vallenari}, {Prusti}, {de Bruijne}, {Babusiaux}, {Bailer-Jones}, {Biermann}, {Evans}, {Eyer}, {Jansen}, {Jordi}, {Klioner}, {Lammers}, {Lindegren}, {Luri}, {Mignard}, {Panem}, {Pourbaix}, {Randich}, {Sartoretti}, {Siddiqui}, {Soubiran}, {van Leeuwen}, {Walton}, {Arenou}, {Bastian}, {Cropper}, {Drimmel}, {Katz}, {Lattanzi}, {Bakker}, {Cacciari}, {Casta{\~n}eda}, {Chaoul}, {Cheek}, {De Angeli}, {Fabricius}, {Guerra}, {Holl}, {Masana}, {Messineo}, {Mowlavi}, {Nienartowicz}, {Panuzzo}, {Portell}, {Riello}, {Seabroke}, {Tanga}, {Th{\'e}venin}, {Gracia-Abril}, {Comoretto}, {Garcia-Reinaldos}, {Teyssier}, {Altmann}, {Andrae}, {Audard}, {Bellas-Velidis}, {Benson}, {Berthier}, {Blomme}, {Burgess}, {Busso}, {Carry}, {Cellino}, {Clementini}, {Clotet}, {Creevey}, {Davidson}, {De Ridder}, {Delchambre}, {Dell'Oro}, {Ducourant}, {Fern{\'a}ndez-Hern{\'a}ndez}, {Fouesneau}, {Fr{\'e}mat}, {Galluccio}, {Garc{\'\i}a-Torres},
  {Gonz{\'a}lez-N{\'u}{\~n}ez}, {Gonz{\'a}lez-Vidal}, {Gosset}, {Guy}, {Halbwachs}, {Hambly}, {Harrison}, {Hern{\'a}ndez}, {Hestroffer}, {Hodgkin}, {Hutton}, {Jasniewicz}, {Jean-Antoine-Piccolo}, {Jordan}, {Korn}, {Krone-Martins}, {Lanzafame}, {Lebzelter}, {L{\"o}ffler}, {Manteiga}, {Marrese}, {Mart{\'\i}n-Fleitas}, {Moitinho}, {Mora}, {Muinonen}, {Osinde}, {Pancino}, {Pauwels}, {Petit}, {Recio-Blanco}, {Richards}, {Rimoldini}, {Robin}, {Sarro}, {Siopis}, {Smith}, {Sozzetti}, {S{\"u}veges}, {Torra}, {van Reeven}, {Abbas}, {Abreu Aramburu}, {Accart}, {Aerts}, {Altavilla}, {{\'A}lvarez}, {Alvarez}, {Alves}, {Anderson}, {Andrei}, {Anglada Varela}, {Antiche}, {Antoja}, {Arcay}, {Astraatmadja}, {Bach}, {Baker}, {Balaguer-N{\'u}{\~n}ez}, {Balm}, {Barache}, {Barata}, {Barbato}, {Barblan}, {Barklem}, {Barrado}, {Barros}, {Barstow}, {Bartholom{\'e} Mu{\~n}oz}, {Bassilana}, {Becciani}, {Bellazzini}, {Berihuete}, {Bertone}, {Bianchi}, {Bienaym{\'e}}, {Blanco-Cuaresma}, {Boch}, {Boeche}, {Bombrun}, {Borrachero},
  {Bossini}, {Bouquillon}, {Bourda}, {Bragaglia}, {Bramante}, {Breddels}, {Bressan}, {Brouillet}, {Br{\"u}semeister}, {Brugaletta}, {Bucciarelli}, {Burlacu}, {Busonero}, {Butkevich}, {Buzzi}, {Caffau}, {Cancelliere}, {Cannizzaro}, {Cantat-Gaudin}, {Carballo}, {Carlucci}, {Carrasco}, {Casamiquela}, {Castellani}, {Castro-Ginard}, {Charlot}, {Chemin}, {Chiavassa}, {Cocozza}, {Costigan}, {Cowell}, {Crifo}, {Crosta}, {Crowley}, {Cuypers}, {Dafonte}, {Damerdji}, {Dapergolas}, {David}, {David}, {de Laverny}, {De Luise}, {De March}, {de Martino}, {de Souza}, {de Torres}, {Debosscher}, {del Pozo}, {Delbo}, {Delgado}, {Delgado}, {Di Matteo}, {Diakite}, {Diener}, {Distefano}, {Dolding}, {Drazinos}, {Dur{\'a}n}, {Edvardsson}, {Enke}, {Eriksson}, {Esquej}, {Eynard Bontemps}, {Fabre}, {Fabrizio}, {Faigler}, {Falc{\~a}o}, {Farr{\`a}s Casas}, {Federici}, {Fedorets}, {Fernique}, {Figueras}, {Filippi}, {Findeisen}, {Fonti}, {Fraile}, {Fraser}, {Fr{\'e}zouls}, {Gai}, {Galleti}, {Garabato}, {Garc{\'\i}a-Sedano}, {Garofalo},
  {Garralda}, {Gavel}, {Gavras}, {Gerssen}, {Geyer}, {Giacobbe}, {Gilmore}, {Girona}, {Giuffrida}, {Glass}, {Gomes}, {Granvik}, {Gueguen}, {Guerrier}, {Guiraud}, {Guti{\'e}rrez-S{\'a}nchez}, {Haigron}, {Hatzidimitriou}, {Hauser}, {Haywood}, {Heiter}, {Helmi}, {Heu}, {Hilger}, {Hobbs}, {Hofmann}, {Holland}, {Huckle}, {Hypki}, {Icardi}, {Jan{\ss}en}, {Jevardat de Fombelle}, {Jonker}, {Juh{\'a}sz}, {Julbe}, {Karampelas}, {Kewley}, {Klar}, {Kochoska}, {Kohley}, {Kolenberg}, {Kontizas}, {Kontizas}, {Koposov}, {Kordopatis}, {Kostrzewa-Rutkowska}, {Koubsky}, {Lambert}, {Lanza}, {Lasne}, {Lavigne}, {Le Fustec}, {Le Poncin-Lafitte}, {Lebreton}, {Leccia}, {Leclerc}, {Lecoeur-Taibi}, {Lenhardt}, {Leroux}, {Liao}, {Licata}, {Lindstr{\o}m}, {Lister}, {Livanou}, {Lobel}, {L{\'o}pez}, {Managau}, {Mann}, {Mantelet}, {Marchal}, {Marchant}, {Marconi}, {Marinoni}, {Marschalk{\'o}}, {Marshall}, {Martino}, {Marton}, {Mary}, {Massari}, {Matijevi{\v{c}}}, {Mazeh}, {McMillan}, {Messina}, {Michalik}, {Millar}, {Molina}, {Molinaro},
  {Moln{\'a}r}, {Montegriffo}, {Mor}, {Morbidelli}, {Morel}, {Morris}, {Mulone}, {Muraveva}, {Musella}, {Nelemans}, {Nicastro}, {Noval}, {O'Mullane}, {Ord{\'e}novic}, {Ord{\'o}{\~n}ez-Blanco}, {Osborne}, {Pagani}, {Pagano}, {Pailler}, {Palacin}, {Palaversa}, {Panahi}, {Pawlak}, {Piersimoni}, {Pineau}, {Plachy}, {Plum}, {Poggio}, {Poujoulet}, {Pr{\v{s}}a}, {Pulone}, {Racero}, {Ragaini}, {Rambaux}, {Ramos-Lerate}, {Regibo}, {Reyl{\'e}}, {Riclet}, {Ripepi}, {Riva}, {Rivard}, {Rixon}, {Roegiers}, {Roelens}, {Romero-G{\'o}mez}, {Rowell}, {Royer}, {Ruiz-Dern}, {Sadowski}, {Sagrist{\`a} Sell{\'e}s}, {Sahlmann}, {Salgado}, {Salguero}, {Sanna}, {Santana-Ros}, {Sarasso}, {Savietto}, {Schultheis}, {Sciacca}, {Segol}, {Segovia}, {S{\'e}gransan}, {Shih}, {Siltala}, {Silva}, {Smart}, {Smith}, {Solano}, {Solitro}, {Sordo}, {Soria Nieto}, {Souchay}, {Spagna}, {Spoto}, {Stampa}, {Steele}, {Steidelm{\"u}ller}, {Stephenson}, {Stoev}, {Suess}, {Surdej}, {Szabados}, {Szegedi-Elek}, {Tapiador}, {Taris}, {Tauran}, {Taylor},
  {Teixeira}, {Terrett}, {Teyssandier}, {Thuillot}, {Titarenko}, {Torra Clotet}, {Turon}, {Ulla}, {Utrilla}, {Uzzi}, {Vaillant}, {Valentini}, {Valette}, {van Elteren}, {Van Hemelryck}, {van Leeuwen}, {Vaschetto}, {Vecchiato}, {Veljanoski}, {Viala}, {Vicente}, {Vogt}, {von Essen}, {Voss}, {Votruba}, {Voutsinas}, {Walmsley}, {Weiler}, {Wertz}, {Wevers}, {Wyrzykowski}, {Yoldas}, {{\v{Z}}erjal}, {Ziaeepour}, {Zorec}, {Zschocke}, {Zucker}, {Zurbach}, \& {Zwitter}}]{Gaiadr2}
{Gaia Collaboration}, {Brown}, A.~G.~A., {Vallenari}, A., {et~al.} 2018, \aap, 616, A1

\bibitem[{{Gaia Collaboration} {et~al.}(2021){Gaia Collaboration}, {Brown}, {Vallenari}, {Prusti}, {de Bruijne}, {Babusiaux}, {Biermann}, {Creevey}, {Evans}, {Eyer}, {Hutton}, {Jansen}, {Jordi}, {Klioner}, {Lammers}, {Lindegren}, {Luri}, {Mignard}, {Panem}, {Pourbaix}, {Randich}, {Sartoretti}, {Soubiran}, {Walton}, {Arenou}, {Bailer-Jones}, {Bastian}, {Cropper}, {Drimmel}, {Katz}, {Lattanzi}, {van Leeuwen}, {Bakker}, {Cacciari}, {Casta{\~n}eda}, {De Angeli}, {Ducourant}, {Fabricius}, {Fouesneau}, {Fr{\'e}mat}, {Guerra}, {Guerrier}, {Guiraud}, {Jean-Antoine Piccolo}, {Masana}, {Messineo}, {Mowlavi}, {Nicolas}, {Nienartowicz}, {Pailler}, {Panuzzo}, {Riclet}, {Roux}, {Seabroke}, {Sordo}, {Tanga}, {Th{\'e}venin}, {Gracia-Abril}, {Portell}, {Teyssier}, {Altmann}, {Andrae}, {Bellas-Velidis}, {Benson}, {Berthier}, {Blomme}, {Brugaletta}, {Burgess}, {Busso}, {Carry}, {Cellino}, {Cheek}, {Clementini}, {Damerdji}, {Davidson}, {Delchambre}, {Dell'Oro}, {Fern{\'a}ndez-Hern{\'a}ndez}, {Galluccio}, {Garc{\'\i}a-Lario},
  {Garcia-Reinaldos}, {Gonz{\'a}lez-N{\'u}{\~n}ez}, {Gosset}, {Haigron}, {Halbwachs}, {Hambly}, {Harrison}, {Hatzidimitriou}, {Heiter}, {Hern{\'a}ndez}, {Hestroffer}, {Hodgkin}, {Holl}, {Jan{\ss}en}, {Jevardat de Fombelle}, {Jordan}, {Krone-Martins}, {Lanzafame}, {L{\"o}ffler}, {Lorca}, {Manteiga}, {Marchal}, {Marrese}, {Moitinho}, {Mora}, {Muinonen}, {Osborne}, {Pancino}, {Pauwels}, {Petit}, {Recio-Blanco}, {Richards}, {Riello}, {Rimoldini}, {Robin}, {Roegiers}, {Rybizki}, {Sarro}, {Siopis}, {Smith}, {Sozzetti}, {Ulla}, {Utrilla}, {van Leeuwen}, {van Reeven}, {Abbas}, {Abreu Aramburu}, {Accart}, {Aerts}, {Aguado}, {Ajaj}, {Altavilla}, {{\'A}lvarez}, {{\'A}lvarez Cid-Fuentes}, {Alves}, {Anderson}, {Anglada Varela}, {Antoja}, {Audard}, {Baines}, {Baker}, {Balaguer-N{\'u}{\~n}ez}, {Balbinot}, {Balog}, {Barache}, {Barbato}, {Barros}, {Barstow}, {Bartolom{\'e}}, {Bassilana}, {Bauchet}, {Baudesson-Stella}, {Becciani}, {Bellazzini}, {Bernet}, {Bertone}, {Bianchi}, {Blanco-Cuaresma}, {Boch}, {Bombrun}, {Bossini},
  {Bouquillon}, {Bragaglia}, {Bramante}, {Breedt}, {Bressan}, {Brouillet}, {Bucciarelli}, {Burlacu}, {Busonero}, {Butkevich}, {Buzzi}, {Caffau}, {Cancelliere}, {C{\'a}novas}, {Cantat-Gaudin}, {Carballo}, {Carlucci}, {Carnerero}, {Carrasco}, {Casamiquela}, {Castellani}, {Castro-Ginard}, {Castro Sampol}, {Chaoul}, {Charlot}, {Chemin}, {Chiavassa}, {Cioni}, {Comoretto}, {Cooper}, {Cornez}, {Cowell}, {Crifo}, {Crosta}, {Crowley}, {Dafonte}, {Dapergolas}, {David}, {David}, {de Laverny}, {De Luise}, {De March}, {De Ridder}, {de Souza}, {de Teodoro}, {de Torres}, {del Peloso}, {del Pozo}, {Delbo}, {Delgado}, {Delgado}, {Delisle}, {Di Matteo}, {Diakite}, {Diener}, {Distefano}, {Dolding}, {Eappachen}, {Edvardsson}, {Enke}, {Esquej}, {Fabre}, {Fabrizio}, {Faigler}, {Fedorets}, {Fernique}, {Fienga}, {Figueras}, {Fouron}, {Fragkoudi}, {Fraile}, {Franke}, {Gai}, {Garabato}, {Garcia-Gutierrez}, {Garc{\'\i}a-Torres}, {Garofalo}, {Gavras}, {Gerlach}, {Geyer}, {Giacobbe}, {Gilmore}, {Girona}, {Giuffrida}, {Gomel}, {Gomez},
  {Gonzalez-Santamaria}, {Gonz{\'a}lez-Vidal}, {Granvik}, {Guti{\'e}rrez-S{\'a}nchez}, {Guy}, {Hauser}, {Haywood}, {Helmi}, {Hidalgo}, {Hilger}, {H{\l}adczuk}, {Hobbs}, {Holland}, {Huckle}, {Jasniewicz}, {Jonker}, {Juaristi Campillo}, {Julbe}, {Karbevska}, {Kervella}, {Khanna}, {Kochoska}, {Kontizas}, {Kordopatis}, {Korn}, {Kostrzewa-Rutkowska}, {Kruszy{\'n}ska}, {Lambert}, {Lanza}, {Lasne}, {Le Campion}, {Le Fustec}, {Lebreton}, {Lebzelter}, {Leccia}, {Leclerc}, {Lecoeur-Taibi}, {Liao}, {Licata}, {Lindstr{\o}m}, {Lister}, {Livanou}, {Lobel}, {Madrero Pardo}, {Managau}, {Mann}, {Marchant}, {Marconi}, {Marcos Santos}, {Marinoni}, {Marocco}, {Marshall}, {Martin Polo}, {Mart{\'\i}n-Fleitas}, {Masip}, {Massari}, {Mastrobuono-Battisti}, {Mazeh}, {McMillan}, {Messina}, {Michalik}, {Millar}, {Mints}, {Molina}, {Molinaro}, {Moln{\'a}r}, {Montegriffo}, {Mor}, {Morbidelli}, {Morel}, {Morris}, {Mulone}, {Munoz}, {Muraveva}, {Murphy}, {Musella}, {Noval}, {Ord{\'e}novic}, {Orr{\`u}}, {Osinde}, {Pagani}, {Pagano},
  {Palaversa}, {Palicio}, {Panahi}, {Pawlak}, {Pe{\~n}alosa Esteller}, {Penttil{\"a}}, {Piersimoni}, {Pineau}, {Plachy}, {Plum}, {Poggio}, {Poretti}, {Poujoulet}, {Pr{\v{s}}a}, {Pulone}, {Racero}, {Ragaini}, {Rainer}, {Raiteri}, {Rambaux}, {Ramos}, {Ramos-Lerate}, {Re Fiorentin}, {Regibo}, {Reyl{\'e}}, {Ripepi}, {Riva}, {Rixon}, {Robichon}, {Robin}, {Roelens}, {Rohrbasser}, {Romero-G{\'o}mez}, {Rowell}, {Royer}, {Rybicki}, {Sadowski}, {Sagrist{\`a} Sell{\'e}s}, {Sahlmann}, {Salgado}, {Salguero}, {Samaras}, {Sanchez Gimenez}, {Sanna}, {Santove{\~n}a}, {Sarasso}, {Schultheis}, {Sciacca}, {Segol}, {Segovia}, {S{\'e}gransan}, {Semeux}, {Shahaf}, {Siddiqui}, {Siebert}, {Siltala}, {Slezak}, {Smart}, {Solano}, {Solitro}, {Souami}, {Souchay}, {Spagna}, {Spoto}, {Steele}, {Steidelm{\"u}ller}, {Stephenson}, {S{\"u}veges}, {Szabados}, {Szegedi-Elek}, {Taris}, {Tauran}, {Taylor}, {Teixeira}, {Thuillot}, {Tonello}, {Torra}, {Torra}, {Turon}, {Unger}, {Vaillant}, {van Dillen}, {Vanel}, {Vecchiato}, {Viala}, {Vicente},
  {Voutsinas}, {Weiler}, {Wevers}, {Wyrzykowski}, {Yoldas}, {Yvard}, {Zhao}, {Zorec}, {Zucker}, {Zurbach}, \& {Zwitter}}]{Gaiaedr3}
{Gaia Collaboration}, {Brown}, A.~G.~A., {Vallenari}, A., {et~al.} 2021, \aap, 649, A1

\bibitem[{{Galleti} {et~al.}(2007){Galleti}, {Bellazzini}, {Federici}, {Buzzoni}, \& {Fusi Pecci}}]{Galleti2007}
{Galleti}, S., {Bellazzini}, M., {Federici}, L., {Buzzoni}, A., \& {Fusi Pecci}, F. 2007, \aap, 471, 127

\bibitem[{{Galleti} {et~al.}(2004{\natexlab{a}}){Galleti}, {Bellazzini}, \& {Ferraro}}]{Galleti2004distance}
{Galleti}, S., {Bellazzini}, M., \& {Ferraro}, F.~R. 2004{\natexlab{a}}, \aap, 423, 925

\bibitem[{{Galleti} {et~al.}(2004{\natexlab{b}}){Galleti}, {Federici}, {Bellazzini}, {Fusi Pecci}, \& {Macrina}}]{Galleti2004}
{Galleti}, S., {Federici}, L., {Bellazzini}, M., {Fusi Pecci}, F., \& {Macrina}, S. 2004{\natexlab{b}}, \aap, 416, 917

\bibitem[{{Gonz{\'a}lez} {et~al.}(2014){Gonz{\'a}lez}, {Kravtsov}, \& {Gnedin}}]{Gonzales2014}
{Gonz{\'a}lez}, R.~E., {Kravtsov}, A.~V., \& {Gnedin}, N.~Y. 2014, \apj, 793, 91

\bibitem[{{Hammer} {et~al.}(2013){Hammer}, {Yang}, {Fouquet}, {Pawlowski}, {Kroupa}, {Puech}, {Flores}, \& {Wang}}]{Hammer2013}
{Hammer}, F., {Yang}, Y., {Fouquet}, S., {et~al.} 2013, \mnras, 431, 3543

\bibitem[{{Hodge} {et~al.}(1999){Hodge}, {Balsley}, {Wyder}, \& {Skelton}}]{Hodge1999}
{Hodge}, P.~W., {Balsley}, J., {Wyder}, T.~K., \& {Skelton}, B.~P. 1999, \pasp, 111, 685

\bibitem[{{Huang} {et~al.}(2015){Huang}, {Liu}, {Yuan}, {Xiang}, {Huo}, {Chen}, {Zhang}, \& {Hou}}]{Huang2015}
{Huang}, Y., {Liu}, X.~W., {Yuan}, H.~B., {et~al.} 2015, \mnras, 449, 162

\bibitem[{{Kahn} \& {Woltjer}(1959)}]{Kahn1959}
{Kahn}, F.~D. \& {Woltjer}, L. 1959, \apj, 130, 705

\bibitem[{{Kam} {et~al.}(2017){Kam}, {Carignan}, {Chemin}, {Foster}, {Elson}, \& {Jarrett}}]{Kam2017}
{Kam}, S.~Z., {Carignan}, C., {Chemin}, L., {et~al.} 2017, \aj, 154, 41

\bibitem[{{Klypin} {et~al.}(2002){Klypin}, {Zhao}, \& {Somerville}}]{Klypin2002}
{Klypin}, A., {Zhao}, H., \& {Somerville}, R.~S. 2002, \apj, 573, 597

\bibitem[{{Liao} {et~al.}(2019){Liao}, {Qi}, {Guo}, \& {Cao}}]{Liao2019}
{Liao}, S.-L., {Qi}, Z.-X., {Guo}, S.-F., \& {Cao}, Z.-H. 2019, Research in Astronomy and Astrophysics, 19, 029

\bibitem[{{Lindegren} {et~al.}(2021{\natexlab{a}}){Lindegren}, {Bastian}, {Biermann}, {Bombrun}, {de Torres}, {Gerlach}, {Geyer}, {Hern{\'a}ndez}, {Hilger}, {Hobbs}, {Klioner}, {Lammers}, {McMillan}, {Ramos-Lerate}, {Steidelm{\"u}ller}, {Stephenson}, \& {van Leeuwen}}]{Lindegren2021par}
{Lindegren}, L., {Bastian}, U., {Biermann}, M., {et~al.} 2021{\natexlab{a}}, \aap, 649, A4

\bibitem[{{Lindegren} {et~al.}(2021{\natexlab{b}}){Lindegren}, {Klioner}, {Hern{\'a}ndez}, {Bombrun}, {Ramos-Lerate}, {Steidelm{\"u}ller}, {Bastian}, {Biermann}, {de Torres}, {Gerlach}, {Geyer}, {Hilger}, {Hobbs}, {Lammers}, {McMillan}, {Stephenson}, {Casta{\~n}eda}, {Davidson}, {Fabricius}, {Gracia-Abril}, {Portell}, {Rowell}, {Teyssier}, {Torra}, {Bartolom{\'e}}, {Clotet}, {Garralda}, {Gonz{\'a}lez-Vidal}, {Torra}, {Abbas}, {Altmann}, {Anglada Varela}, {Balaguer-N{\'u}{\~n}ez}, {Balog}, {Barache}, {Becciani}, {Bernet}, {Bertone}, {Bianchi}, {Bouquillon}, {Brown}, {Bucciarelli}, {Busonero}, {Butkevich}, {Buzzi}, {Cancelliere}, {Carlucci}, {Charlot}, {Cioni}, {Crosta}, {Crowley}, {del Peloso}, {del Pozo}, {Drimmel}, {Esquej}, {Fienga}, {Fraile}, {Gai}, {Garcia-Reinaldos}, {Guerra}, {Hambly}, {Hauser}, {Jan{\ss}en}, {Jordan}, {Kostrzewa-Rutkowska}, {Lattanzi}, {Liao}, {Licata}, {Lister}, {L{\"o}ffler}, {Marchant}, {Masip}, {Mignard}, {Mints}, {Molina}, {Mora}, {Morbidelli}, {Murphy}, {Pagani}, {Panuzzo},
  {Pe{\~n}alosa Esteller}, {Poggio}, {Re Fiorentin}, {Riva}, {Sagrist{\`a} Sell{\'e}s}, {Sanchez Gimenez}, {Sarasso}, {Sciacca}, {Siddiqui}, {Smart}, {Souami}, {Spagna}, {Steele}, {Taris}, {Utrilla}, {van Reeven}, \& {Vecchiato}}]{Lindegren2021ast}
{Lindegren}, L., {Klioner}, S.~A., {Hern{\'a}ndez}, J., {et~al.} 2021{\natexlab{b}}, \aap, 649, A2

\bibitem[{{Lynden-Bell}(1981)}]{Lynden-Bell1981}
{Lynden-Bell}, D. 1981, The Observatory, 101, 111

\bibitem[{{Massey} \& {Evans}(2016)}]{Massey2016rsg}
{Massey}, P. \& {Evans}, K.~A. 2016, \apj, 826, 224

\bibitem[{{Massey} {et~al.}(2021){Massey}, {Neugent}, {Levesque}, {Drout}, \& {Courteau}}]{Massey2021}
{Massey}, P., {Neugent}, K.~F., {Levesque}, E.~M., {Drout}, M.~R., \& {Courteau}, S. 2021, \aj, 161, 79

\bibitem[{{Massey} {et~al.}(2016){Massey}, {Neugent}, \& {Smart}}]{Massey2016}
{Massey}, P., {Neugent}, K.~F., \& {Smart}, B.~M. 2016, \aj, 152, 62

\bibitem[{{Massey} \& {Olsen}(2003)}]{Massey2003}
{Massey}, P. \& {Olsen}, K.~A.~G. 2003, \aj, 126, 2867

\bibitem[{{Massey} {et~al.}(2006){Massey}, {Olsen}, {Hodge}, {Strong}, {Jacoby}, {Schlingman}, \& {Smith}}]{Massey2006}
{Massey}, P., {Olsen}, K.~A.~G., {Hodge}, P.~W., {et~al.} 2006, \aj, 131, 2478

\bibitem[{{Massey} {et~al.}(2009){Massey}, {Silva}, {Levesque}, {Plez}, {Olsen}, {Clayton}, {Meynet}, \& {Maeder}}]{Massey2009}
{Massey}, P., {Silva}, D.~R., {Levesque}, E.~M., {et~al.} 2009, \apj, 703, 420

\bibitem[{{McConnachie}(2012)}]{McConnachie2012AJ}
{McConnachie}, A.~W. 2012, \aj, 144, 4

\bibitem[{{McConnachie} {et~al.}(2006){McConnachie}, {Chapman}, {Ibata}, {Ferguson}, {Irwin}, {Lewis}, {Tanvir}, \& {Martin}}]{McConnachie2006}
{McConnachie}, A.~W., {Chapman}, S.~C., {Ibata}, R.~A., {et~al.} 2006, \apjl, 647, L25

\bibitem[{{McConnachie} {et~al.}(2009){McConnachie}, {Irwin}, {Ibata}, {Dubinski}, {Widrow}, {Martin}, {C{\^o}t{\'e}}, {Dotter}, {Navarro}, {Ferguson}, {Puzia}, {Lewis}, {Babul}, {Barmby}, {Bienaym{\'e}}, {Chapman}, {Cockcroft}, {Collins}, {Fardal}, {Harris}, {Huxor}, {Mackey}, {Pe{\~n}arrubia}, {Rich}, {Richer}, {Siebert}, {Tanvir}, {Valls-Gabaud}, \& {Venn}}]{McConnachie2009}
{McConnachie}, A.~W., {Irwin}, M.~J., {Ibata}, R.~A., {et~al.} 2009, \nat, 461, 66

\bibitem[{{Merrett} {et~al.}(2006){Merrett}, {Merrifield}, {Douglas}, {Kuijken}, {Romanowsky}, {Napolitano}, {Arnaboldi}, {Capaccioli}, {Freeman}, {Gerhard}, {Coccato}, {Carter}, {Evans}, {Wilkinson}, {Halliday}, \& {Bridges}}]{Merrett2006}
{Merrett}, H.~R., {Merrifield}, M.~R., {Douglas}, N.~G., {et~al.} 2006, \mnras, 369, 120

\bibitem[{{Patel} {et~al.}(2017){Patel}, {Besla}, \& {Sohn}}]{Patel2017}
{Patel}, E., {Besla}, G., \& {Sohn}, S.~T. 2017, \mnras, 464, 3825

\bibitem[{{Rastorguev} {et~al.}(2017){Rastorguev}, {Utkin}, {Zabolotskikh}, {Dambis}, {Bajkova}, \& {Bobylev}}]{Rastorguev2017}
{Rastorguev}, A.~S., {Utkin}, N.~D., {Zabolotskikh}, M.~V., {et~al.} 2017, Astrophysical Bulletin, 72, 122

\bibitem[{{Riello} {et~al.}(2021){Riello}, {De Angeli}, {Evans}, {Montegriffo}, {Carrasco}, {Busso}, {Palaversa}, {Burgess}, {Diener}, {Davidson}, {Rowell}, {Fabricius}, {Jordi}, {Bellazzini}, {Pancino}, {Harrison}, {Cacciari}, {van Leeuwen}, {Hambly}, {Hodgkin}, {Osborne}, {Altavilla}, {Barstow}, {Brown}, {Castellani}, {Cowell}, {De Luise}, {Gilmore}, {Giuffrida}, {Hidalgo}, {Holland}, {Marinoni}, {Pagani}, {Piersimoni}, {Pulone}, {Ragaini}, {Rainer}, {Richards}, {Sanna}, {Walton}, {Weiler}, \& {Yoldas}}]{Riello2021}
{Riello}, M., {De Angeli}, F., {Evans}, D.~W., {et~al.} 2021, \aap, 649, A3

\bibitem[{{Rubin} \& {Ford}(1970)}]{Rubin1970}
{Rubin}, V.~C. \& {Ford}, W.~Kent, J. 1970, \apj, 159, 379

\bibitem[{{Rusterucci} {et~al.}(2024){Rusterucci}, {Martin}, {Starkenburg}, \& {Ibata}}]{Rusterucci2024}
{Rusterucci}, S., {Martin}, N.~F., {Starkenburg}, E., \& {Ibata}, R. 2024, \aap, 692, A30

\bibitem[{{Salomon} {et~al.}(2021){Salomon}, {Ibata}, {Reyl{\'e}}, {Famaey}, {Libeskind}, {McConnachie}, \& {Hoffman}}]{Salomon2021}
{Salomon}, J.~B., {Ibata}, R., {Reyl{\'e}}, C., {et~al.} 2021, \mnras, 507, 2592

\bibitem[{{Salomon} {et~al.}(2016){Salomon}, {Ibata}, {Famaey}, {Martin}, \& {Lewis}}]{Salomon2016}
{Salomon}, J.~B., {Ibata}, R.~A., {Famaey}, B., {Martin}, N.~F., \& {Lewis}, G.~F. 2016, \mnras, 456, 4432

\bibitem[{{Sanders} {et~al.}(2012){Sanders}, {Caldwell}, {McDowell}, \& {Harding}}]{Sanders2012}
{Sanders}, N.~E., {Caldwell}, N., {McDowell}, J., \& {Harding}, P. 2012, \apj, 758, 133

\bibitem[{{Sarajedini} \& {Mancone}(2007)}]{Sarajedini2007}
{Sarajedini}, A. \& {Mancone}, C.~L. 2007, \aj, 134, 447

\bibitem[{{Sawala} {et~al.}(2024){Sawala}, {Delhomelle}, {Deason}, {Frenk}, {Johansson}, {Keitaanranta}, {Rawlings}, \& {Wright}}]{Sawala2024}
{Sawala}, T., {Delhomelle}, J., {Deason}, A.~J., {et~al.} 2024, arXiv e-prints, arXiv:2408.00064

\bibitem[{{Sawala} {et~al.}(2023){Sawala}, {Teeriaho}, \& {Johansson}}]{Sawala2023}
{Sawala}, T., {Teeriaho}, M., \& {Johansson}, P.~H. 2023, \mnras, 521, 4863

\bibitem[{{Schiavi} {et~al.}(2020){Schiavi}, {Capuzzo-Dolcetta}, {Arca-Sedda}, \& {Spera}}]{Schiavi2020}
{Schiavi}, R., {Capuzzo-Dolcetta}, R., {Arca-Sedda}, M., \& {Spera}, M. 2020, \aap, 642, A30

\bibitem[{{Sohn} {et~al.}(2012){Sohn}, {Anderson}, \& {van der Marel}}]{Sohn2012}
{Sohn}, S.~T., {Anderson}, J., \& {van der Marel}, R.~P. 2012, \apj, 753, 7

\bibitem[{{Stanek} \& {Garnavich}(1998)}]{Stanek1998}
{Stanek}, K.~Z. \& {Garnavich}, P.~M. 1998, \apjl, 503, L131

\bibitem[{{Tolstoy} {et~al.}(2009){Tolstoy}, {Hill}, \& {Tosi}}]{Tolstoy2009}
{Tolstoy}, E., {Hill}, V., \& {Tosi}, M. 2009, \araa, 47, 371

\bibitem[{van~den Bergh(2000)}]{van2000}
van~den Bergh, S. 2000, The galaxies of the Local Group

\bibitem[{{van der Marel} {et~al.}(2012{\natexlab{a}}){van der Marel}, {Besla}, {Cox}, {Sohn}, \& {Anderson}}]{vdm2012b}
{van der Marel}, R.~P., {Besla}, G., {Cox}, T.~J., {Sohn}, S.~T., \& {Anderson}, J. 2012{\natexlab{a}}, \apj, 753, 9

\bibitem[{{van der Marel} {et~al.}(2012{\natexlab{b}}){van der Marel}, {Fardal}, {Besla}, {Beaton}, {Sohn}, {Anderson}, {Brown}, \& {Guhathakurta}}]{vdm2012a}
{van der Marel}, R.~P., {Fardal}, M., {Besla}, G., {et~al.} 2012{\natexlab{b}}, \apj, 753, 8

\bibitem[{{van der Marel} {et~al.}(2019){van der Marel}, {Fardal}, {Sohn}, {Patel}, {Besla}, {del Pino}, {Sahlmann}, \& {Watkins}}]{vdm2019}
{van der Marel}, R.~P., {Fardal}, M.~A., {Sohn}, S.~T., {et~al.} 2019, \apj, 872, 24

\bibitem[{{van der Marel} \& {Guhathakurta}(2008)}]{vdm2008}
{van der Marel}, R.~P. \& {Guhathakurta}, P. 2008, \apj, 678, 187

\bibitem[{{Walterbos} \& {Kennicutt}(1988)}]{Walterbos1988}
{Walterbos}, R.~A.~M. \& {Kennicutt}, Jr., R.~C. 1988, \aap, 198, 61

\bibitem[{{Watkins} {et~al.}(2013){Watkins}, {Evans}, \& {van de Ven}}]{Watkins2013}
{Watkins}, L.~L., {Evans}, N.~W., \& {van de Ven}, G. 2013, \mnras, 430, 971

\bibitem[{{Weisz} {et~al.}(2014){Weisz}, {Dolphin}, {Skillman}, {Holtzman}, {Gilbert}, {Dalcanton}, \& {Williams}}]{Weisz2014}
{Weisz}, D.~R., {Dolphin}, A.~E., {Skillman}, E.~D., {et~al.} 2014, \apj, 789, 147

\bibitem[{{Wu} {et~al.}(2025){Wu}, {Huang}, {Zhang}, {Yuan}, {Huo}, \& {Liu}}]{Wu2024}
{Wu}, H., {Huang}, Y., {Zhang}, H., {et~al.} 2025, Research in Astronomy and Astrophysics, 25, 015012

\bibitem[{{Zhang} {et~al.}(2020){Zhang}, {Chen}, {Huo}, {Zhang}, {Xiang}, {Yuan}, {Huang}, {Wang}, \& {Liu}}]{Zhang2020}
{Zhang}, M., {Chen}, B.-Q., {Huo}, Z.-Y., {et~al.} 2020, Research in Astronomy and Astrophysics, 20, 097

\bibitem[{{Zhang} {et~al.}(2024){Zhang}, {Chen}, {Chen}, {Sun}, \& {Tian}}]{Zhang2024}
{Zhang}, X., {Chen}, B., {Chen}, P., {Sun}, J., \& {Tian}, Z. 2024, \mnras, 528, 2653

\bibitem[{{Zhou} {et~al.}(2023){Zhou}, {Li}, {Huang}, \& {Zhang}}]{Zhou2023}
{Zhou}, Y., {Li}, X., {Huang}, Y., \& {Zhang}, H. 2023, \apj, 946, 73

\end{thebibliography}
\end{document}